\documentclass[copyright]{eptcs}

\usepackage{times}
\usepackage{amsmath}
\usepackage{amssymb}

\usepackage{proof}
\usepackage{prooftree}
\usepackage{moreverb}
\usepackage{color}
\usepackage{ifthen}
\usepackage{supertabular}
\usepackage{xspace}
\usepackage{prooftree}

\begin{document}

\newtheorem{thm}{Theorem}
\newtheorem{lemma}[thm]{Lemma}
\newtheorem{corollary}[thm]{Corollary}
\newtheorem{definition}[thm]{Definition}
\newtheorem{remark}[thm]{Remark}
\newtheorem{proposition}[thm]{Proposition}
\newtheorem{notn}[thm]{Notation}
\newtheorem{observation}[thm]{Observation}

\newcommand{\Teqt}[0]{\texorpdfstring{\ensuremath{\texttt{T}^{\texttt{eq}\downarrow}}\xspace}{Teqt}}
\newcommand{\Tdept}[0]{\texttt{T}^{\texttt{dep}\downarrow}}
\newcommand{\interp}[1]{[\negthinspace[#1]\negthinspace]}

\newcommand{\alt}{\ensuremath{\ \mid\ }}	
\newcommand{\defeq}{\ensuremath{\stackrel{\mathrm{def}}{=}}}
\newcommand{\pfcase}[2]{ \subsection{Case: $\ottdrulename{#1}$ } \[ #2 \] \linebreak \noindent}

\newcommand{\AS}[1]{
\textbf{\color{red}{AS: #1}}}
\newcommand{\scw}[1]{
\textbf{\color{red}{SW: #1}}}
\newcommand{\vs}[1]{
\textbf{\color{red}{VS: #1}}}

\newif \ifTR \TRfalse

\ifTR
\title{Termination Casts: A Flexible Approach to Termination with General Recursion (Technical Appendix)}
\else
\title{Termination Casts: A Flexible Approach to Termination with General Recursion}
\fi

\author{Aaron Stump \\
\institute{Computer Science\\
The University of Iowa}
\email{astump@acm.org}
\and
Vilhelm Sj\"oberg\\
\institute{Computer and Information Science\\
University of Pennsylvania}
\email{vilhelm@cis.upenn.edu}
\and
Stephanie Weirich \\
\institute{Computer and Information Science\\
University of Pennsylvania}
\email{sweirich@cis.upenn.edu}
}

\def\authorrunning{Stump, Sj\"oberg, and Weirich}
\def\titlerunning{Termination Casts}

\newcommand{\ottdrule}[4][]{{\displaystyle\frac{\begin{array}{l}#2\end{array}}{#3}\quad\ottdrulename{#4}}}
\newcommand{\ottusedrule}[1]{\[#1\]}
\newcommand{\ottpremise}[1]{ #1 \\}
\newenvironment{ottdefnblock}[3][]{ \framebox{\mbox{#2}} \quad #3 \\[0pt]}{}
\newenvironment{ottfundefnblock}[3][]{ \framebox{\mbox{#2}} \quad #3 \\[0pt]\begin{displaymath}\begin{array}{l}}{\end{array}\end{displaymath}}
\newcommand{\ottfunclause}[2]{ #1 \equiv #2 \\}
\newcommand{\ottnt}[1]{\mathit{#1}}
\newcommand{\ottmv}[1]{\mathit{#1}}
\newcommand{\ottkw}[1]{\mathbf{#1}}
\newcommand{\ottcom}[1]{\text{#1}}
\newcommand{\ottdrulename}[1]{\textsc{#1}}
\newcommand{\ottcomplu}[5]{\overline{#1}^{\,#2\in #3 #4 #5}}
\newcommand{\ottcompu}[3]{\overline{#1}^{\,#2<#3}}
\newcommand{\ottcomp}[2]{\overline{#1}^{\,#2}}
\newcommand{\ottgrammartabular}[1]{\begin{supertabular}{llcllllll}#1\end{supertabular}}
\newcommand{\ottmetavartabular}[1]{\begin{supertabular}{ll}#1\end{supertabular}}
\newcommand{\ottrulehead}[3]{$#1$ & & $#2$ & & & \multicolumn{2}{l}{#3}}
\newcommand{\ottprodline}[6]{& & $#1$ & $#2$ & $#3 #4$ & $#5$ & $#6$}
\newcommand{\ottinterrule}{\\[5.0mm]}
\newcommand{\ottafterlastrule}{\\}
\newcommand{\ottmetavars}{
\ottmetavartabular{
 $ \mathit{termvar} ,\, \mathit{x} ,\, \mathit{f} ,\, \mathit{h} ,\, \mathit{p} ,\, \mathit{q} ,\, \mathit{u} ,\, \mathit{z} $ &  \\
 $ \mathit{avar} ,\, x ,\, f ,\, p ,\, q $ &  \\
 $ \ottmv{index} ,\, \ottmv{i} ,\, \ottmv{n} $ &  \\
}}

\newcommand{\otttyp}{
\ottrulehead{\ottnt{typ}  ,\ \ottnt{T}}{::=}{\ottcom{Types}}\\ 
\ottprodline{|}{\ottkw{nat}}{}{}{}{}\\ 
\ottprodline{|}{ \Pi ^{ \theta }  \mathit{x} \!:\! \ottnt{T} .  \ottnt{T'} }{}{\textsf{bind}\; \mathit{x}\; \textsf{in}\; \ottnt{T'}}{}{}\\ 
\ottprodline{|}{\ottnt{t} \, = \, \ottnt{t'}}{}{}{}{}\\ 
\ottprodline{|}{\ottkw{Terminates}\ \, \ottnt{t}}{}{}{}{}}

\newcommand{\otttheta}{
\ottrulehead{\theta  ,\ \rho}{::=}{\ottcom{Effects}}\\ 
\ottprodline{|}{ \downarrow }{}{}{}{}\\ 
\ottprodline{|}{ ? }{}{}{}{}}

\newcommand{\ottterm}{
\ottrulehead{\ottnt{term}  ,\ \ottnt{t}}{::=}{\ottcom{unannotated terms}}\\ 
\ottprodline{|}{\mathit{x}}{}{}{}{}\\ 
\ottprodline{|}{\ottnt{t} \, \ottnt{t'}}{}{}{}{}\\ 
\ottprodline{|}{\lambda \, \mathit{x} \, . \, \ottnt{t}}{}{\textsf{bind}\; \mathit{x}\; \textsf{in}\; \ottnt{t}}{}{}\\ 
\ottprodline{|}{0}{}{}{}{}\\ 
\ottprodline{|}{\ottkw{Suc} \, \ottnt{t}}{}{}{}{}\\ 
\ottprodline{|}{\ottkw{join}}{}{}{}{}\\ 
\ottprodline{|}{\ottkw{terminates}}{}{}{}{}\\ 
\ottprodline{|}{\ottkw{contra}}{}{}{}{}\\ 
\ottprodline{|}{\ottkw{abort}}{}{}{}{}\\ 
\ottprodline{|}{ \ottkw{rec} \  \mathit{f} ( \mathit{x} ) =  \ottnt{t} }{}{\textsf{bind}\; \mathit{x}\; \textsf{in}\; \ottnt{t}}{}{}\\ 
\ottprodline{|}{ \ottkw{case} \  \ottnt{t} \  \ottnt{t'} \  \ottnt{t''} }{}{}{}{}}

\newcommand{\ottatyp}{
\ottrulehead{\ottnt{atyp}  ,\ \ottnt{S}}{::=}{\ottcom{Annotated language types}}\\ 
\ottprodline{|}{\ottkw{nat}}{}{}{}{}\\ 
\ottprodline{|}{ \Pi ^{ \theta }  x \!:\! \ottnt{S} .  \ottnt{S'} }{}{\textsf{bind}\; x\; \textsf{in}\; \ottnt{S'}}{}{}\\ 
\ottprodline{|}{\ottnt{a} \, = \, \ottnt{a'}}{}{}{}{}\\ 
\ottprodline{|}{\ottkw{Terminates}\ \, \ottnt{a}}{}{}{}{}}

\newcommand{\ottaterm}{
\ottrulehead{\ottnt{aterm}  ,\ \ottnt{a}}{::=}{\ottcom{annotated terms}}\\ 
\ottprodline{|}{x}{}{}{}{}\\ 
\ottprodline{|}{\ottnt{a} \, \ottnt{a'}}{}{}{}{}\\ 
\ottprodline{|}{ \lambda^{\textcolor{black}{ \theta } }  x  \textcolor{black}{\!:\!  \ottnt{S} }. \ottnt{a} }{}{\textsf{bind}\; x\; \textsf{in}\; \ottnt{a}}{}{}\\ 
\ottprodline{|}{0}{}{}{}{}\\ 
\ottprodline{|}{\ottkw{Suc} \, \ottnt{a}}{}{}{}{}\\ 
\ottprodline{|}{ \ottkw{join} \ \textcolor{black}{ \ottnt{a} \  \ottnt{a'} } }{}{}{}{}\\ 
\ottprodline{|}{ \textcolor{black}{ \ottkw{conv} \  x . \ottnt{S} }\  \ottnt{a'} \ \textcolor{black}{ \ottnt{a} } }{}{\textsf{bind}\; x\; \textsf{in}\; \ottnt{S}}{}{}\\ 
\ottprodline{|}{ \textcolor{black}{ \ottkw{reflect} }\  \ottnt{a} \ \textcolor{black}{ \ottnt{a'} } }{}{}{}{}\\ 
\ottprodline{|}{ \ottkw{terminates} \ \textcolor{black}{ \ottnt{a} } }{}{}{}{}\\ 
\ottprodline{|}{ \textcolor{black}{ \ottkw{inv} }\  \ottnt{a} \ \textcolor{black}{ \ottnt{a'} } }{}{}{}{}\\ 
\ottprodline{|}{ \ottkw{contra} \ \textcolor{black}{  \ottnt{S} \  \ottnt{a}  } }{}{}{}{}\\ 
\ottprodline{|}{ \ottkw{abort} \ \textcolor{black}{  \ottnt{S} } }{}{}{}{}\\ 
\ottprodline{|}{ \ottkw{rec} _{\textcolor{black}{ \ottkw{nat} } }\  f ( x \  p ) \textcolor{black}{ :  \ottnt{S}  } =  \ottnt{a} }{}{\textsf{bind}\; x\; \textsf{in}\; \ottnt{a}}{}{}\\ 
\ottprodline{|}{ \ottkw{rec} \  f ( x  \textcolor{black}{\!:\! \ottnt{S} }) \textcolor{black}{: \ottnt{S'} } =  \ottnt{a} }{}{\textsf{bind}\; x\; \textsf{in}\; \ottnt{a}}{}{}\\ 
\ottprodline{|}{ \ottkw{case} \ \textcolor{black}{ x  .  \ottnt{S} }\  \ottnt{a} \  \ottnt{a'} \  \ottnt{a''} }{}{\textsf{bind}\; x\; \textsf{in}\; \ottnt{S}}{}{}}

\newcommand{\ottstp}{
\ottrulehead{\ottnt{stp}  ,\ \ottnt{A}}{::=}{\ottcom{Simple Types}}\\ 
\ottprodline{|}{\ottkw{nat}}{}{}{}{}\\ 
\ottprodline{|}{ \ottnt{A}  \to  \ottnt{A'} }{}{}{}{}}

\newcommand{\ottform}{
\ottrulehead{\ottnt{form}  ,\ \ottnt{F}}{::=}{\ottcom{Formulas}}\\ 
\ottprodline{|}{\ottkw{True}}{}{}{}{}\\ 
\ottprodline{|}{ \forall  \mathit{x}  :  \ottnt{A}  .  \ottnt{F} }{}{}{}{}\\ 
\ottprodline{|}{ \ottnt{F}  \Rightarrow  \ottnt{F'} }{}{}{}{}\\ 
\ottprodline{|}{ \ottnt{F}  \wedge  \ottnt{F'} }{}{}{}{}\\ 
\ottprodline{|}{\ottkw{Terminates}\ \, \ottnt{t}}{}{}{}{}\\ 
\ottprodline{|}{\ottnt{t} \, = \, \ottnt{t'}}{}{}{}{}}

\newcommand{\ottenv}{
\ottrulehead{\ottnt{env}  ,\ \Gamma}{::=}{}\\ 
\ottprodline{|}{ \cdot }{}{}{}{}\\ 
\ottprodline{|}{\Gamma \, , \, \mathit{x} \, : \, \ottnt{T}}{}{}{}{}\\ 
\ottprodline{|}{\Gamma \, , \, x \, : \, \ottnt{S}}{}{}{}{}}

\newcommand{\ottST}{
\ottrulehead{\Sigma}{::=}{}\\ 
\ottprodline{|}{ \cdot }{}{}{}{}\\ 
\ottprodline{|}{\Sigma \, , \, \mathit{x} \, : \, \ottnt{A}}{}{}{}{}}

\newcommand{\ottH}{
\ottrulehead{\ottnt{H}}{::=}{}\\ 
\ottprodline{|}{ \cdot }{}{}{}{}\\ 
\ottprodline{|}{\ottnt{H} \, , \, \ottnt{F}}{}{}{}{}}

\newcommand{\ottD}{
\ottrulehead{\Delta}{::=}{}\\ 
\ottprodline{|}{ \cdot }{}{}{}{}\\ 
\ottprodline{|}{\Delta \, , \, \mathit{x} \, : \, \ottnt{F}}{}{}{}{}}

\newcommand{\ottC}{
\ottrulehead{{\cal C}}{::=}{}\\ 
\ottprodline{|}{\ottkw{\_\_}}{}{}{}{}\\ 
\ottprodline{|}{\ottkw{Suc} \, {\cal C}}{}{}{}{}\\ 
\ottprodline{|}{{\cal C} \, \ottnt{t}}{}{}{}{}\\ 
\ottprodline{|}{\ottnt{v} \, {\cal C}}{}{}{}{}\\ 
\ottprodline{|}{\ottkw{case} \, {\cal C} \, \ottnt{t} \, \ottnt{t'}}{}{}{}{}}

\newcommand{\ottCa}{
\ottrulehead{{\cal C}_{a}}{::=}{}\\ 
\ottprodline{|}{\ottkw{\_\_}}{}{}{}{}\\ 
\ottprodline{|}{\ottkw{Suc} \, {\cal C}_{a}}{}{}{}{}\\ 
\ottprodline{|}{{\cal C}_{a} \, \ottnt{a}}{}{}{}{}\\ 
\ottprodline{|}{\ottnt{a} \, {\cal C}_{a}}{}{}{}{}\\ 
\ottprodline{|}{\ottkw{case} \, x \, . \, \ottnt{S} \, {\cal C}_{a} \, \ottnt{a} \, \ottnt{a'}}{}{}{}{}\\ 
\ottprodline{|}{\ottkw{conv} \, x \, . \, \ottnt{S} \, {\cal C}_{a} \, \ottnt{a}}{}{}{}{}\\ 
\ottprodline{|}{\ottkw{reflect} \, {\cal C}_{a} \, \ottnt{a}}{}{}{}{}}

\newcommand{\ottv}{
\ottrulehead{\ottnt{v}}{::=}{}\\ 
\ottprodline{|}{\mathit{x}}{}{}{}{}\\ 
\ottprodline{|}{0}{}{}{}{}\\ 
\ottprodline{|}{\ottkw{Suc} \, \ottnt{v}}{}{}{}{}\\ 
\ottprodline{|}{\lambda \, \mathit{x} \, . \, \ottnt{t}}{}{}{}{}\\ 
\ottprodline{|}{ \ottkw{rec} \  \mathit{f} ( \mathit{x} ) =  \ottnt{t} }{}{}{}{}\\ 
\ottprodline{|}{\ottkw{join}}{}{}{}{}\\ 
\ottprodline{|}{\ottkw{terminates}}{}{}{}{}\\ 
\ottprodline{|}{\ottkw{contra}}{}{}{}{}}

\newcommand{\ottgrammar}{\ottgrammartabular{
\otttyp\ottinterrule
\otttheta\ottinterrule
\ottterm\ottinterrule
\ottatyp\ottinterrule
\ottaterm\ottinterrule
\ottstp\ottinterrule
\ottform\ottinterrule
\ottenv\ottinterrule
\ottST\ottinterrule
\ottH\ottinterrule
\ottD\ottinterrule
\ottC\ottinterrule
\ottCa\ottinterrule
\ottv\ottafterlastrule
}}

\newcommand{\ottdruleRedXXCtxt}[1]{\ottdrule[#1]{%
\ottpremise{\ottnt{t} \, \leadsto_\beta \, \ottnt{t'}}%
}{
{\cal C} \, [ \, \ottnt{t} \, ] \, \leadsto \, {\cal C} \, [ \, \ottnt{t'} \, ]}{%
{\ottdrulename{Red\_Ctxt}}{}%
}}

\newcommand{\ottdruleRedXXAbort}[1]{\ottdrule[#1]{%
}{
{\cal C} \, [ \, \ottkw{abort} \, ] \, \leadsto \, \ottkw{abort}}{%
{\ottdrulename{Red\_Abort}}{}%
}}

\newcommand{\ottdefnred}[1]{\begin{ottdefnblock}[#1]{$\ottnt{t} \, \leadsto \, \ottnt{t'}$}{}
\ottusedrule{\ottdruleRedXXCtxt{}}
\ottusedrule{\ottdruleRedXXAbort{}}
\end{ottdefnblock}}

\newcommand{\ottdruleBetaXXAppAbs}[1]{\ottdrule[#1]{%
}{
( \, \lambda \, \mathit{x} \, . \, \ottnt{t} \, ) \, \ottnt{v} \, \leadsto_\beta \, [ \, \ottnt{v} \, / \, \mathit{x} \, ] \, \ottnt{t}}{%
{\ottdrulename{Beta\_AppAbs}}{}%
}}

\newcommand{\ottdruleBetaXXCaseZero}[1]{\ottdrule[#1]{%
}{
 \ottkw{case} \  0 \  \ottnt{t} \  \ottnt{t'}  \, \leadsto_\beta \, \ottnt{t}}{%
{\ottdrulename{Beta\_CaseZero}}{}%
}}

\newcommand{\ottdruleBetaXXCaseSuc}[1]{\ottdrule[#1]{%
}{
 \ottkw{case} \  ( \, \ottkw{Suc} \, \ottnt{v} \, ) \  \ottnt{t} \  \ottnt{t'}  \, \leadsto_\beta \, \ottnt{t'} \, \ottnt{v}}{%
{\ottdrulename{Beta\_CaseSuc}}{}%
}}

\newcommand{\ottdruleBetaXXAppRec}[1]{\ottdrule[#1]{%
}{
( \,  \ottkw{rec} \  \mathit{f} ( \mathit{x} ) =  \ottnt{t}  \, ) \, \ottnt{v} \, \leadsto_\beta \, [ \, \ottnt{v} \, / \, \mathit{x} \, ] \, [ \,  \ottkw{rec} \  \mathit{f} ( \mathit{x} ) =  \ottnt{t}  \, / \, \mathit{f} \, ] \, \ottnt{t}}{%
{\ottdrulename{Beta\_AppRec}}{}%
}}

\newcommand{\ottdefnbeta}[1]{\begin{ottdefnblock}[#1]{$\ottnt{t} \, \leadsto_\beta \, \ottnt{t'}$}{}
\ottusedrule{\ottdruleBetaXXAppAbs{}}
\ottusedrule{\ottdruleBetaXXCaseZero{}}
\ottusedrule{\ottdruleBetaXXCaseSuc{}}
\ottusedrule{\ottdruleBetaXXAppRec{}}
\end{ottdefnblock}}

\newcommand{\ottdruleOkXXempty}[1]{\ottdrule[#1]{%
}{
 \cdot  \, \vdash \, \ottkw{Ok}}{%
{\ottdrulename{Ok\_empty}}{}%
}}

\newcommand{\ottdruleOkXXcons}[1]{\ottdrule[#1]{%
\ottpremise{ \Gamma \, \vdash \, \ottkw{Ok} \ \ \ \ \   \Gamma  \vdash  \ottnt{T}  }%
}{
\Gamma \, , \, \mathit{x} \, : \, \ottnt{T} \, \vdash \, \ottkw{Ok}}{%
{\ottdrulename{Ok\_cons}}{}%
}}

\newcommand{\ottdefnctxwf}[1]{\begin{ottdefnblock}[#1]{$\Gamma \, \vdash \, \ottkw{Ok}$}{\ottcom{ctx well-formedness}}
\ottusedrule{\ottdruleOkXXempty{}}
\ottusedrule{\ottdruleOkXXcons{}}
\end{ottdefnblock}}

\newcommand{\ottdruleSubXXAx}[1]{\ottdrule[#1]{%
}{
 \downarrow  \, \leq \,  ? }{%
{\ottdrulename{Sub\_Ax}}{}%
}}

\newcommand{\ottdruleSubXXRefl}[1]{\ottdrule[#1]{%
}{
\theta \, \leq \, \theta}{%
{\ottdrulename{Sub\_Refl}}{}%
}}

\newcommand{\ottdefnsubeff}[1]{\begin{ottdefnblock}[#1]{$\theta \, \leq \, \theta'$}{\ottcom{subeffect relation}}
\ottusedrule{\ottdruleSubXXAx{}}
\ottusedrule{\ottdruleSubXXRefl{}}
\end{ottdefnblock}}

\newcommand{\ottdrulekXXNat}[1]{\ottdrule[#1]{%
\ottpremise{\Gamma \, \vdash \, \ottkw{Ok}}%
}{
 \Gamma  \vdash  \ottkw{nat} }{%
{\ottdrulename{k\_Nat}}{}%
}}

\newcommand{\ottdrulekXXPi}[1]{\ottdrule[#1]{%
\ottpremise{ \Gamma \, , \, \mathit{x} \, : \, \ottnt{T}  \vdash  \ottnt{T'} }%
}{
 \Gamma  \vdash   \Pi ^{ \theta }  \mathit{x} \!:\! \ottnt{T} .  \ottnt{T'}  }{%
{\ottdrulename{k\_Pi}}{}%
}}

\newcommand{\ottdrulekXXEq}[1]{\ottdrule[#1]{%
\ottpremise{  \Gamma  \vdash  \ottnt{t}  :  \ottnt{T} \   ?   \ \ \ \ \   \Gamma  \vdash  \ottnt{t'}  :  \ottnt{T'} \   ?   }%
}{
 \Gamma  \vdash  \ottnt{t} \, = \, \ottnt{t'} }{%
{\ottdrulename{k\_Eq}}{}%
}}

\newcommand{\ottdrulekXXTerm}[1]{\ottdrule[#1]{%
\ottpremise{ \Gamma  \vdash  \ottnt{t}  :  \ottnt{T} \   ?  }%
}{
 \Gamma  \vdash  \ottkw{Terminates}\ \, \ottnt{t} }{%
{\ottdrulename{k\_Term}}{}%
}}

\newcommand{\ottdefnkassn}[1]{\begin{ottdefnblock}[#1]{$ \Gamma  \vdash  \ottnt{T} $}{\ottcom{Type well-formedness}}
\ottusedrule{\ottdrulekXXNat{}}
\ottusedrule{\ottdrulekXXPi{}}
\ottusedrule{\ottdrulekXXEq{}}
\ottusedrule{\ottdrulekXXTerm{}}
\end{ottdefnblock}}

\newcommand{\ottdruleTXXVar}[1]{\ottdrule[#1]{%
\ottpremise{  \Gamma ( \mathit{x} ) =  \ottnt{T}  \ \ \ \ \  \Gamma \, \vdash \, \ottkw{Ok} }%
}{
 \Gamma  \vdash  \mathit{x}  :  \ottnt{T} \  \theta }{%
{\ottdrulename{T\_Var}}{}%
}}

\newcommand{\ottdruleTXXJoin}[1]{\ottdrule[#1]{%
\ottpremise{  \ottnt{t}  \leadsto^*  \ottnt{t_{{\mathrm{0}}}}  \ \ \ \ \   \ottnt{t'}  \leadsto^*  \ottnt{t_{{\mathrm{0}}}}  }%
\ottpremise{  \Gamma  \vdash  \ottnt{t}  :  \ottnt{T} \   ?   \ \ \ \ \   \Gamma  \vdash  \ottnt{t'}  :  \ottnt{T'} \   ?   }%
}{
 \Gamma  \vdash  \ottkw{join}  :  \ottnt{t} \, = \, \ottnt{t'} \  \theta }{%
{\ottdrulename{T\_Join}}{}%
}}

\newcommand{\ottdruleTXXConv}[1]{\ottdrule[#1]{%
\ottpremise{ \Gamma  \vdash  \ottnt{t}  :  [ \, \ottnt{t_{{\mathrm{2}}}} \, / \, \mathit{x} \, ] \, \ottnt{T} \  \theta }%
\ottpremise{  \Gamma  \vdash  \ottnt{t'}  :  \ottnt{t_{{\mathrm{1}}}} \, = \, \ottnt{t_{{\mathrm{2}}}} \   \downarrow   \ \ \ \ \   \Gamma  \vdash  [ \, \ottnt{t_{{\mathrm{1}}}} \, / \, \mathit{x} \, ] \, \ottnt{T}  }%
}{
 \Gamma  \vdash  \ottnt{t}  :  [ \, \ottnt{t_{{\mathrm{1}}}} \, / \, \mathit{x} \, ] \, \ottnt{T} \  \theta }{%
{\ottdrulename{T\_Conv}}{}%
}}

\newcommand{\ottdruleTXXReflect}[1]{\ottdrule[#1]{%
\ottpremise{ \Gamma  \vdash  \ottnt{t}  :  \ottnt{T} \   ?  }%
\ottpremise{ \Gamma  \vdash  \ottnt{t'}  :  \ottkw{Terminates}\ \, \ottnt{t} \   \downarrow  }%
}{
 \Gamma  \vdash  \ottnt{t}  :  \ottnt{T} \  \theta }{%
{\ottdrulename{T\_Reflect}}{}%
}}

\newcommand{\ottdruleTXXReify}[1]{\ottdrule[#1]{%
\ottpremise{ \Gamma  \vdash  \ottnt{t}  :  \ottnt{T} \   \downarrow  }%
}{
 \Gamma  \vdash  \ottkw{terminates}  :  \ottkw{Terminates}\ \, \ottnt{t} \  \theta }{%
{\ottdrulename{T\_Reify}}{}%
}}

\newcommand{\ottdruleTXXCtxTerm}[1]{\ottdrule[#1]{%
\ottpremise{ \Gamma  \vdash  \ottnt{t}  :  \ottkw{Terminates}\ \, {\cal C} \, [ \, \ottnt{t'} \, ] \  \theta }%
}{
 \Gamma  \vdash  \ottnt{t}  :  \ottkw{Terminates}\ \, \ottnt{t'} \  \theta }{%
{\ottdrulename{T\_CtxTerm}}{}%
}}

\newcommand{\ottdruleTXXAbs}[1]{\ottdrule[#1]{%
\ottpremise{  \Gamma \, , \, \mathit{x} \, : \, \ottnt{T'}  \vdash  \ottnt{t}  :  \ottnt{T} \  \rho  \ \ \ \ \   \Gamma  \vdash   \Pi ^{ \rho }  \mathit{x} \!:\! \ottnt{T'} .  \ottnt{T}   }%
}{
 \Gamma  \vdash  \lambda \, \mathit{x} \, . \, \ottnt{t}  :   \Pi ^{ \rho }  \mathit{x} \!:\! \ottnt{T'} .  \ottnt{T}  \  \theta }{%
{\ottdrulename{T\_Abs}}{}%
}}

\newcommand{\ottdruleTXXApp}[1]{\ottdrule[#1]{%
\ottpremise{  \Gamma  \vdash  \ottnt{t}  :   \Pi ^{ \rho }  \mathit{x} \!:\! \ottnt{T'} .  \ottnt{T}  \  \theta  \ \ \ \ \     \Gamma  \vdash  \ottnt{t'}  :  \ottnt{T'} \  \theta  \ \ \ \ \  \rho \, \leq \, \theta   }%
}{
 \Gamma  \vdash  \ottnt{t} \, \ottnt{t'}  :  [ \, \ottnt{t'} \, / \, \mathit{x} \, ] \, \ottnt{T} \  \theta }{%
{\ottdrulename{T\_App}}{}%
}}

\newcommand{\ottdruleTXXZero}[1]{\ottdrule[#1]{%
\ottpremise{\Gamma \, \vdash \, \ottkw{Ok}}%
}{
 \Gamma  \vdash  0  :  \ottkw{nat} \  \theta }{%
{\ottdrulename{T\_Zero}}{}%
}}

\newcommand{\ottdruleTXXSuc}[1]{\ottdrule[#1]{%
\ottpremise{ \Gamma  \vdash  \ottnt{t}  :  \ottkw{nat} \  \theta }%
}{
 \Gamma  \vdash  \ottkw{Suc} \, \ottnt{t}  :  \ottkw{nat} \  \theta }{%
{\ottdrulename{T\_Suc}}{}%
}}

\newcommand{\ottdruleTXXRec}[1]{\ottdrule[#1]{%
\ottpremise{ \Gamma \, , \, \mathit{f} \, : \,  \Pi ^{  ?  }  \mathit{x} \!:\! \ottnt{T'} .  \ottnt{T}  \, , \, \mathit{x} \, : \, \ottnt{T'}  \vdash  \ottnt{t}  :  \ottnt{T} \   ?  }%
}{
 \Gamma  \vdash   \ottkw{rec} \  \mathit{f} ( \mathit{x} ) =  \ottnt{t}   :   \Pi ^{  ?  }  \mathit{x} \!:\! \ottnt{T'} .  \ottnt{T}  \  \theta }{%
{\ottdrulename{T\_Rec}}{}%
}}

\newcommand{\ottdruleTXXRecNat}[1]{\ottdrule[#1]{%
\ottpremise{\mathit{p} \, \not\in \, \ottkw{fv} \, \ottnt{t}}%
\ottpremise{ \Gamma \, , \, \mathit{f} \, : \,  \Pi ^{  ?  }  \mathit{x} \!:\! \ottkw{nat} .  \ottnt{T}  \, , \, \mathit{x} \, : \, \ottkw{nat} \, , \, \mathit{p} \, : \,  \Pi ^{  \downarrow  }  \mathit{x_{{\mathrm{1}}}} \!:\! \ottkw{nat} .   \Pi ^{  \downarrow  }  \mathit{p'} \!:\! \mathit{x} \, = \, \ottkw{Suc} \, \mathit{x_{{\mathrm{1}}}} .  \ottkw{Terminates}\ \, ( \, \mathit{f} \, \mathit{x_{{\mathrm{1}}}} \, )    \vdash  \ottnt{t}  :  \ottnt{T} \   \downarrow  }%
}{
 \Gamma  \vdash   \ottkw{rec} \  \mathit{f} ( \mathit{x} ) =  \ottnt{t}   :   \Pi ^{  \downarrow  }  \mathit{x} \!:\! \ottkw{nat} .  \ottnt{T}  \  \theta }{%
{\ottdrulename{T\_RecNat}}{}%
}}

\newcommand{\ottdruleTXXCase}[1]{\ottdrule[#1]{%
\ottpremise{  \Gamma  \vdash  \ottnt{t}  :  \ottkw{nat} \  \theta  \ \ \ \ \   \Gamma  \vdash  \ottnt{t'}  :  [ \, 0 \, / \, \mathit{x} \, ] \, \ottnt{T} \  \theta  }%
\ottpremise{  \Gamma  \vdash  \ottnt{t''}  :   \Pi ^{ \rho }  \mathit{x'} \!:\! \ottkw{nat} .  [ \, \ottkw{Suc} \, \mathit{x'} \, / \, \mathit{x} \, ] \, \ottnt{T}  \  \theta  \ \ \ \ \  \rho \, \leq \, \theta }%
}{
 \Gamma  \vdash   \ottkw{case} \  \ottnt{t} \  \ottnt{t'} \  \ottnt{t''}   :  [ \, \ottnt{t} \, / \, \mathit{x} \, ] \, \ottnt{T} \  \theta }{%
{\ottdrulename{T\_Case}}{}%
}}

\newcommand{\ottdruleTXXContra}[1]{\ottdrule[#1]{%
\ottpremise{ \Gamma  \vdash  \ottnt{t}  :  0 \, = \, \ottkw{Suc} \, \ottnt{t'} \   \downarrow  }%
}{
 \Gamma  \vdash  \ottkw{contra}  :  \ottnt{T} \  \theta }{%
{\ottdrulename{T\_Contra}}{}%
}}

\newcommand{\ottdruleTXXAbort}[1]{\ottdrule[#1]{%
\ottpremise{\Gamma \, \vdash \, \ottkw{Ok}}%
}{
 \Gamma  \vdash  \ottkw{abort}  :  \ottnt{T} \   ?  }{%
{\ottdrulename{T\_Abort}}{}%
}}

\newcommand{\ottdefntyassn}[1]{\begin{ottdefnblock}[#1]{$ \Gamma  \vdash  \ottnt{t}  :  \ottnt{T} \  \theta $}{\ottcom{Type assignment}}
\ottusedrule{\ottdruleTXXVar{}}
\ottusedrule{\ottdruleTXXJoin{}}
\ottusedrule{\ottdruleTXXConv{}}
\ottusedrule{\ottdruleTXXReflect{}}
\ottusedrule{\ottdruleTXXReify{}}
\ottusedrule{\ottdruleTXXCtxTerm{}}
\ottusedrule{\ottdruleTXXAbs{}}
\ottusedrule{\ottdruleTXXApp{}}
\ottusedrule{\ottdruleTXXZero{}}
\ottusedrule{\ottdruleTXXSuc{}}
\ottusedrule{\ottdruleTXXRec{}}
\ottusedrule{\ottdruleTXXRecNat{}}
\ottusedrule{\ottdruleTXXCase{}}
\ottusedrule{\ottdruleTXXContra{}}
\ottusedrule{\ottdruleTXXAbort{}}
\end{ottdefnblock}}

\newcommand{\ottdruleOkaXXempty}[1]{\ottdrule[#1]{%
}{
  \cdot   \Vdash  \ottkw{Ok} }{%
{\ottdrulename{Oka\_empty}}{}%
}}

\newcommand{\ottdruleOkaXXcons}[1]{\ottdrule[#1]{%
\ottpremise{  \Gamma  \Vdash  \ottkw{Ok}  \ \ \ \ \   \Gamma  \Vdash  \ottnt{S}  }%
}{
 \Gamma \, , \, x \, : \, \ottnt{S}  \Vdash  \ottkw{Ok} }{%
{\ottdrulename{Oka\_cons}}{}%
}}

\newcommand{\ottdefnctxwfa}[1]{\begin{ottdefnblock}[#1]{$ \Gamma  \Vdash  \ottkw{Ok} $}{\ottcom{ctx well-formedness}}
\ottusedrule{\ottdruleOkaXXempty{}}
\ottusedrule{\ottdruleOkaXXcons{}}
\end{ottdefnblock}}

\newcommand{\ottdruleSXXNat}[1]{\ottdrule[#1]{%
\ottpremise{ \Gamma  \Vdash  \ottkw{Ok} }%
}{
 \Gamma  \Vdash  \ottkw{nat} }{%
{\ottdrulename{S\_Nat}}{}%
}}

\newcommand{\ottdruleSXXPi}[1]{\ottdrule[#1]{%
\ottpremise{  \Gamma  \Vdash  \ottnt{S}  \ \ \ \ \   \Gamma \, , \, x \, : \, \ottnt{S}  \Vdash  \ottnt{S'}  }%
}{
 \Gamma  \Vdash   \Pi ^{ \theta }  x \!:\! \ottnt{S} .  \ottnt{S'}  }{%
{\ottdrulename{S\_Pi}}{}%
}}

\newcommand{\ottdruleSXXEq}[1]{\ottdrule[#1]{%
\ottpremise{  \Gamma  \Vdash  \ottnt{a}  :  \ottnt{S} \   ?   \ \ \ \ \   \Gamma  \Vdash  \ottnt{a'}  :  \ottnt{S'} \   ?   }%
\ottpremise{  \Gamma  \Vdash  \ottnt{S}  \ \ \ \ \   \Gamma  \Vdash  \ottnt{S'}  }%
}{
 \Gamma  \Vdash  \ottnt{a} \, = \, \ottnt{a'} }{%
{\ottdrulename{S\_Eq}}{}%
}}

\newcommand{\ottdruleSXXTerm}[1]{\ottdrule[#1]{%
\ottpremise{ \Gamma  \Vdash  \ottnt{a}  :  \ottnt{S} \   ?  }%
}{
 \Gamma  \Vdash  \ottkw{Terminates}\ \, \ottnt{a} }{%
{\ottdrulename{S\_Term}}{}%
}}

\newcommand{\ottdefnkassna}[1]{\begin{ottdefnblock}[#1]{$ \Gamma  \Vdash  \ottnt{S} $}{\ottcom{Type well-formedness}}
\ottusedrule{\ottdruleSXXNat{}}
\ottusedrule{\ottdruleSXXPi{}}
\ottusedrule{\ottdruleSXXEq{}}
\ottusedrule{\ottdruleSXXTerm{}}
\end{ottdefnblock}}

\newcommand{\ottdruleAXXVar}[1]{\ottdrule[#1]{%
\ottpremise{  \Gamma ( x ) =  \ottkw{T}  \ \ \ \ \   \Gamma  \Vdash  \ottkw{Ok}  }%
}{
 \Gamma  \Vdash  x  :  \ottnt{S} \  \theta }{%
{\ottdrulename{A\_Var}}{}%
}}

\newcommand{\ottdruleAXXJoin}[1]{\ottdrule[#1]{%
\ottpremise{  \ottkw{\mbox{$\mid$}} \, \ottnt{a} \, \ottkw{\mbox{$\mid$}}  \leadsto^N  \ottnt{t}  \ \ \ \ \   \ottkw{\mbox{$\mid$}} \, \ottnt{a'} \, \ottkw{\mbox{$\mid$}}  \leadsto^N  \ottnt{t}  }%
\ottpremise{  \Gamma  \Vdash  \ottnt{a}  :  \ottnt{S} \   ?   \ \ \ \ \   \Gamma  \Vdash  \ottnt{a'}  :  \ottnt{S'} \   ?   }%
}{
 \Gamma  \Vdash   \ottkw{join} \ \textcolor{black}{ \ottnt{a} \  \ottnt{a'} }   :  \ottnt{a} \, = \, \ottnt{a'} \  \theta }{%
{\ottdrulename{A\_Join}}{}%
}}

\newcommand{\ottdruleAXXConv}[1]{\ottdrule[#1]{%
\ottpremise{ \Gamma  \Vdash  \ottnt{a}  :  [ \, \ottnt{a_{{\mathrm{2}}}} \, / \, x \, ] \, \ottnt{S} \  \theta }%
\ottpremise{  \Gamma  \Vdash  \ottnt{a'}  :  \ottnt{a_{{\mathrm{1}}}} \, = \, \ottnt{a_{{\mathrm{2}}}} \   \downarrow   \ \ \ \ \   \Gamma  \Vdash  [ \, \ottnt{a_{{\mathrm{1}}}} \, / \, x \, ] \, \ottnt{S}  }%
}{
 \Gamma  \Vdash   \textcolor{black}{ \ottkw{conv} \  x . \ottnt{S} }\  \ottnt{a} \ \textcolor{black}{ \ottnt{a'} }   :  [ \, \ottnt{a_{{\mathrm{1}}}} \, / \, x \, ] \, \ottnt{S} \  \theta }{%
{\ottdrulename{A\_Conv}}{}%
}}

\newcommand{\ottdruleAXXReflect}[1]{\ottdrule[#1]{%
\ottpremise{  \Gamma  \Vdash  \ottnt{a}  :  \ottnt{S} \   ?   \ \ \ \ \   \Gamma  \Vdash  \ottnt{a'}  :  \ottkw{Terminates}\ \, \ottnt{a} \   \downarrow   }%
}{
 \Gamma  \Vdash   \textcolor{black}{ \ottkw{reflect} }\  \ottnt{a} \ \textcolor{black}{ \ottnt{a'} }   :  \ottnt{S} \  \theta }{%
{\ottdrulename{A\_Reflect}}{}%
}}

\newcommand{\ottdruleAXXReify}[1]{\ottdrule[#1]{%
\ottpremise{ \Gamma  \Vdash  \ottnt{a}  :  \ottnt{S} \   \downarrow  }%
}{
 \Gamma  \Vdash   \ottkw{terminates} \ \textcolor{black}{ \ottnt{a} }   :  \ottkw{Terminates}\ \, \ottnt{a} \  \theta }{%
{\ottdrulename{A\_Reify}}{}%
}}

\newcommand{\ottdruleAXXCtxTerm}[1]{\ottdrule[#1]{%
\ottpremise{ \Gamma  \Vdash  \ottnt{a}  :  \ottkw{Terminates}\ \, \ottnt{a''} \  \theta }%
\ottpremise{\ottkw{\mbox{$\mid$}} \, \ottnt{a''} \, \ottkw{\mbox{$\mid$}} \, = \, {\cal C} \, [ \, \ottkw{\mbox{$\mid$}} \, \ottnt{a'} \, \ottkw{\mbox{$\mid$}} \, ]}%
}{
 \Gamma  \Vdash   \textcolor{black}{ \ottkw{inv} }\  \ottnt{a} \ \textcolor{black}{ \ottnt{a'} }   :  \ottkw{Terminates}\ \, \ottnt{a'} \  \theta }{%
{\ottdrulename{A\_CtxTerm}}{}%
}}

\newcommand{\ottdruleAXXAbs}[1]{\ottdrule[#1]{%
\ottpremise{  \Gamma \, , \, x \, : \, \ottnt{S'}  \Vdash  \ottnt{a}  :  \ottnt{S} \  \rho  \ \ \ \ \   \Gamma  \Vdash   \Pi ^{ \rho }  x \!:\! \ottnt{S'} .  \ottnt{S}   }%
}{
 \Gamma  \Vdash   \lambda^{\textcolor{black}{ \rho } }  x  \textcolor{black}{\!:\!  \ottnt{S'} }. \ottnt{a}   :   \Pi ^{ \rho }  x \!:\! \ottnt{S'} .  \ottnt{S}  \  \theta }{%
{\ottdrulename{A\_Abs}}{}%
}}

\newcommand{\ottdruleAXXApp}[1]{\ottdrule[#1]{%
\ottpremise{  \Gamma  \Vdash  \ottnt{a}  :   \Pi ^{ \rho }  x \!:\! \ottnt{S'} .  \ottnt{S}  \  \theta  \ \ \ \ \     \Gamma  \Vdash  \ottnt{a'}  :  \ottnt{S'} \  \theta  \ \ \ \ \  \rho \, \leq \, \theta   }%
}{
 \Gamma  \Vdash  \ottnt{a} \, \ottnt{a'}  :  [ \, \ottnt{a'} \, / \, x \, ] \, \ottnt{S} \  \theta }{%
{\ottdrulename{A\_App}}{}%
}}

\newcommand{\ottdruleAXXZero}[1]{\ottdrule[#1]{%
\ottpremise{ \Gamma  \Vdash  \ottkw{Ok} }%
}{
 \Gamma  \Vdash  0  :  \ottkw{nat} \  \theta }{%
{\ottdrulename{A\_Zero}}{}%
}}

\newcommand{\ottdruleAXXSuc}[1]{\ottdrule[#1]{%
\ottpremise{ \Gamma  \Vdash  \ottnt{a}  :  \ottkw{nat} \  \theta }%
}{
 \Gamma  \Vdash  \ottkw{Suc} \, \ottnt{a}  :  \ottkw{nat} \  \theta }{%
{\ottdrulename{A\_Suc}}{}%
}}

\newcommand{\ottdruleAXXRec}[1]{\ottdrule[#1]{%
\ottpremise{ \Gamma \, , \, f \, : \,  \Pi ^{  ?  }  x \!:\! \ottnt{S'} .  \ottnt{S}  \, , \, x \, : \, \ottnt{S'}  \Vdash  \ottnt{a}  :  \ottnt{S} \   ?  }%
}{
 \Gamma  \Vdash   \ottkw{rec} \  f ( x  \textcolor{black}{\!:\! \ottnt{S'} }) \textcolor{black}{: \ottnt{S} } =  \ottnt{a}   :   \Pi ^{  ?  }  x \!:\! \ottnt{S'} .  \ottnt{S}  \  \theta }{%
{\ottdrulename{A\_Rec}}{}%
}}

\newcommand{\ottdruleAXXRecNat}[1]{\ottdrule[#1]{%
\ottpremise{p \, \not\in \, \ottkw{fv} \, \ottnt{a}}%
\ottpremise{ \Gamma \, , \, f \, : \,  \Pi ^{  ?  }  x \!:\! \ottkw{nat} .  \ottnt{S}  \, , \, x \, : \, \ottkw{nat} \, , \, p \, : \,  \Pi ^{  \downarrow  }  x_{{\mathrm{1}}} \!:\! \ottkw{nat} .   \Pi ^{  \downarrow  }  p' \!:\! x \, = \, \ottkw{Suc} \, x_{{\mathrm{1}}} .  \ottkw{Terminates}\ \, ( \, f \, x_{{\mathrm{1}}} \, )    \Vdash  \ottnt{a}  :  \ottnt{S} \   \downarrow  }%
}{
 \Gamma  \Vdash   \ottkw{rec} _{\textcolor{black}{ \ottkw{nat} } }\  f ( x \  p ) \textcolor{black}{ :  \ottnt{S}  } =  \ottnt{a}   :   \Pi ^{  \downarrow  }  x \!:\! \ottkw{nat} .  \ottnt{S}  \  \theta }{%
{\ottdrulename{A\_RecNat}}{}%
}}

\newcommand{\ottdruleAXXCase}[1]{\ottdrule[#1]{%
\ottpremise{  \Gamma  \Vdash  \ottnt{a}  :  \ottkw{nat} \  \theta  \ \ \ \ \   \Gamma  \Vdash  \ottnt{a'}  :  [ \, 0 \, / \, x \, ] \, \ottnt{S} \  \theta  }%
\ottpremise{ \Gamma  \Vdash  \ottnt{a''}  :   \Pi ^{ \rho }  x' \!:\! \ottkw{nat} .  [ \, \ottkw{Suc} \, x' \, / \, x \, ] \, \ottnt{S}  \  \theta }%
\ottpremise{\rho \, \leq \, \theta}%
}{
 \Gamma  \Vdash   \ottkw{case} \ \textcolor{black}{ x  .  \ottnt{S} }\  \ottnt{a} \  \ottnt{a'} \  \ottnt{a''}   :  [ \, \ottnt{a} \, / \, x \, ] \, \ottnt{S} \  \theta }{%
{\ottdrulename{A\_Case}}{}%
}}

\newcommand{\ottdruleAXXContra}[1]{\ottdrule[#1]{%
\ottpremise{ \Gamma  \Vdash  \ottnt{a}  :  0 \, = \, \ottkw{Suc} \, \ottnt{a'} \   \downarrow  }%
}{
 \Gamma  \Vdash   \ottkw{contra} \ \textcolor{black}{  \ottnt{S} \  \ottnt{a}  }   :  \ottnt{S} \  \theta }{%
{\ottdrulename{A\_Contra}}{}%
}}

\newcommand{\ottdruleAXXAbort}[1]{\ottdrule[#1]{%
\ottpremise{ \Gamma  \Vdash  \ottkw{Ok} }%
}{
 \Gamma  \Vdash   \ottkw{abort} \ \textcolor{black}{  \ottnt{S} }   :  \ottnt{S} \   ?  }{%
{\ottdrulename{A\_Abort}}{}%
}}

\newcommand{\ottdefntyassna}[1]{\begin{ottdefnblock}[#1]{$ \Gamma  \Vdash  \ottnt{a}  :  \ottnt{S} \  \theta $}{}
\ottusedrule{\ottdruleAXXVar{}}
\ottusedrule{\ottdruleAXXJoin{}}
\ottusedrule{\ottdruleAXXConv{}}
\ottusedrule{\ottdruleAXXReflect{}}
\ottusedrule{\ottdruleAXXReify{}}
\ottusedrule{\ottdruleAXXCtxTerm{}}
\ottusedrule{\ottdruleAXXAbs{}}
\ottusedrule{\ottdruleAXXApp{}}
\ottusedrule{\ottdruleAXXZero{}}
\ottusedrule{\ottdruleAXXSuc{}}
\ottusedrule{\ottdruleAXXRec{}}
\ottusedrule{\ottdruleAXXRecNat{}}
\ottusedrule{\ottdruleAXXCase{}}
\ottusedrule{\ottdruleAXXContra{}}
\ottusedrule{\ottdruleAXXAbort{}}
\end{ottdefnblock}}

\newcommand{\ottdruleSTyXXVar}[1]{\ottdrule[#1]{%
\ottpremise{ \Sigma ( \mathit{x} ) =  \ottnt{A} }%
}{
 \Sigma  \vdash  \mathit{x}  :  \ottnt{A} }{%
{\ottdrulename{STy\_Var}}{}%
}}

\newcommand{\ottdruleSTyXXAbs}[1]{\ottdrule[#1]{%
\ottpremise{ \Sigma \, , \, \mathit{x} \, : \, \ottnt{A_{{\mathrm{1}}}}  \vdash  \ottnt{t}  :  \ottnt{A_{{\mathrm{2}}}} }%
}{
 \Sigma  \vdash  \lambda \, \mathit{x} \, . \, \ottnt{t}  :   \ottnt{A_{{\mathrm{1}}}}  \to  \ottnt{A_{{\mathrm{2}}}}  }{%
{\ottdrulename{STy\_Abs}}{}%
}}

\newcommand{\ottdruleSTyXXApp}[1]{\ottdrule[#1]{%
\ottpremise{  \Sigma  \vdash  \ottnt{t_{{\mathrm{1}}}}  :   \ottnt{A_{{\mathrm{2}}}}  \to  \ottnt{A_{{\mathrm{1}}}}   \ \ \ \ \   \Sigma  \vdash  \ottnt{t_{{\mathrm{2}}}}  :  \ottnt{A_{{\mathrm{2}}}}  }%
}{
 \Sigma  \vdash  \ottnt{t_{{\mathrm{1}}}} \, \ottnt{t_{{\mathrm{2}}}}  :  \ottnt{A_{{\mathrm{1}}}} }{%
{\ottdrulename{STy\_App}}{}%
}}

\newcommand{\ottdruleSTyXXZero}[1]{\ottdrule[#1]{%
}{
 \Sigma  \vdash  0  :  \ottkw{nat} }{%
{\ottdrulename{STy\_Zero}}{}%
}}

\newcommand{\ottdruleSTyXXSuc}[1]{\ottdrule[#1]{%
\ottpremise{ \Sigma  \vdash  \ottnt{t}  :  \ottkw{nat} }%
}{
 \Sigma  \vdash  \ottkw{Suc} \, \ottnt{t}  :  \ottkw{nat} }{%
{\ottdrulename{STy\_Suc}}{}%
}}

\newcommand{\ottdruleSTyXXRec}[1]{\ottdrule[#1]{%
\ottpremise{ \Sigma \, , \, \mathit{f} \, : \,  \ottkw{nat}  \to  \ottnt{A}  \, , \, \mathit{x} \, : \, \ottkw{nat}  \vdash  \ottnt{t}  :  \ottnt{A} }%
}{
 \Sigma  \vdash   \ottkw{rec} \  \mathit{f} ( \mathit{x} ) =  \ottnt{t}   :   \ottkw{nat}  \to  \ottnt{A}  }{%
{\ottdrulename{STy\_Rec}}{}%
}}

\newcommand{\ottdruleSTyXXCase}[1]{\ottdrule[#1]{%
\ottpremise{  \Sigma  \vdash  \ottnt{t}  :  \ottkw{nat}  \ \ \ \ \     \Sigma  \vdash  \ottnt{t'}  :  \ottnt{A}  \ \ \ \ \   \Sigma  \vdash  \ottnt{t''}  :   \ottkw{nat}  \to  \ottnt{A}     }%
}{
 \Sigma  \vdash   \ottkw{case} \  \ottnt{t} \  \ottnt{t'} \  \ottnt{t''}   :  \ottnt{A} }{%
{\ottdrulename{STy\_Case}}{}%
}}

\newcommand{\ottdruleSTyXXAbort}[1]{\ottdrule[#1]{%
}{
 \Sigma  \vdash  \ottkw{abort}  :  \ottnt{A} }{%
{\ottdrulename{STy\_Abort}}{}%
}}

\newcommand{\ottdefnstassn}[1]{\begin{ottdefnblock}[#1]{$ \Sigma  \vdash  \ottnt{t}  :  \ottnt{A} $}{\ottcom{Simple-type assignment}}
\ottusedrule{\ottdruleSTyXXVar{}}
\ottusedrule{\ottdruleSTyXXAbs{}}
\ottusedrule{\ottdruleSTyXXApp{}}
\ottusedrule{\ottdruleSTyXXZero{}}
\ottusedrule{\ottdruleSTyXXSuc{}}
\ottusedrule{\ottdruleSTyXXRec{}}
\ottusedrule{\ottdruleSTyXXCase{}}
\ottusedrule{\ottdruleSTyXXAbort{}}
\end{ottdefnblock}}

\newcommand{\ottdrulePvXXAssume}[1]{\ottdrule[#1]{%
\ottpremise{\ottnt{F} \, \in \, \ottnt{H}}%
}{
\Sigma \, ; \, \ottnt{H} \, \vdash \, \ottnt{F}}{%
{\ottdrulename{Pv\_Assume}}{}%
}}

\newcommand{\ottdrulePvXXAlli}[1]{\ottdrule[#1]{%
\ottpremise{ \Sigma \, , \, \mathit{x} \, : \, \ottnt{A} \, ; \, \ottnt{H} \, \vdash \, \ottnt{F} \ \ \ \ \  \mathit{x} \, \not\in \, \ottkw{fv} \, \ottnt{H} }%
}{
\Sigma \, ; \, \ottnt{H} \, \vdash \,  \forall  \mathit{x}  :  \ottnt{A}  .  \ottnt{F} }{%
{\ottdrulename{Pv\_Alli}}{}%
}}

\newcommand{\ottdrulePvXXAlle}[1]{\ottdrule[#1]{%
\ottpremise{ \Sigma \, ; \, \ottnt{H} \, \vdash \,  \forall  \mathit{x}  :  \ottnt{A}  .  \ottnt{F}  \ \ \ \ \   \Sigma  \vdash  \ottnt{t}  :  \ottnt{A}  }%
}{
\Sigma \, ; \, \ottnt{H} \, \vdash \, [ \, \ottnt{t} \, / \, \mathit{x} \, ] \, \ottnt{F}}{%
{\ottdrulename{Pv\_Alle}}{}%
}}

\newcommand{\ottdrulePvXXImpi}[1]{\ottdrule[#1]{%
\ottpremise{\Sigma \, ; \, \ottnt{H} \, , \, \ottnt{F} \, \vdash \, \ottnt{F'}}%
}{
\Sigma \, ; \, \ottnt{H} \, \vdash \,  \ottnt{F}  \Rightarrow  \ottnt{F'} }{%
{\ottdrulename{Pv\_Impi}}{}%
}}

\newcommand{\ottdrulePvXXImpe}[1]{\ottdrule[#1]{%
\ottpremise{ \Sigma \, ; \, \ottnt{H} \, \vdash \,  \ottnt{F}  \Rightarrow  \ottnt{F'}  \ \ \ \ \  \Sigma \, ; \, \ottnt{H} \, \vdash \, \ottnt{F} }%
}{
\Sigma \, ; \, \ottnt{H} \, \vdash \, \ottnt{F'}}{%
{\ottdrulename{Pv\_Impe}}{}%
}}

\newcommand{\ottdrulePvXXAndi}[1]{\ottdrule[#1]{%
\ottpremise{ \Sigma \, ; \, \ottnt{H} \, \vdash \, \ottnt{F} \ \ \ \ \  \Sigma \, ; \, \ottnt{H} \, \vdash \, \ottnt{F'} }%
}{
\Sigma \, ; \, \ottnt{H} \, \vdash \,  \ottnt{F}  \wedge  \ottnt{F'} }{%
{\ottdrulename{Pv\_Andi}}{}%
}}

\newcommand{\ottdrulePvXXAndeOne}[1]{\ottdrule[#1]{%
\ottpremise{\Sigma \, ; \, \ottnt{H} \, \vdash \,  \ottnt{F}  \wedge  \ottnt{F'} }%
}{
\Sigma \, ; \, \ottnt{H} \, \vdash \, \ottnt{F}}{%
{\ottdrulename{Pv\_Ande1}}{}%
}}

\newcommand{\ottdrulePvXXAndeTwo}[1]{\ottdrule[#1]{%
\ottpremise{\Sigma \, ; \, \ottnt{H} \, \vdash \,  \ottnt{F}  \wedge  \ottnt{F'} }%
}{
\Sigma \, ; \, \ottnt{H} \, \vdash \, \ottnt{F'}}{%
{\ottdrulename{Pv\_Ande2}}{}%
}}

\newcommand{\ottdrulePvXXTruei}[1]{\ottdrule[#1]{%
}{
\Sigma \, ; \, \ottnt{H} \, \vdash \, \ottkw{True}}{%
{\ottdrulename{Pv\_Truei}}{}%
}}

\newcommand{\ottdrulePvXXContra}[1]{\ottdrule[#1]{%
\ottpremise{\Sigma \, ; \, \ottnt{H} \, \vdash \, 0 \, = \, \ottkw{Suc} \, \ottnt{t}}%
}{
\Sigma \, ; \, \ottnt{H} \, \vdash \, \ottnt{F}}{%
{\ottdrulename{Pv\_Contra}}{}%
}}

\newcommand{\ottdrulePvXXInd}[1]{\ottdrule[#1]{%
\ottpremise{ \Sigma \, ; \, \ottnt{H} \, \vdash \, [ \, 0 \, / \, \mathit{x} \, ] \, \ottnt{F} \ \ \ \ \  \Sigma \, , \, \mathit{x'} \, : \, \ottkw{nat} \, ; \, \ottnt{H} \, , \, \ottkw{Terminates}\ \, \mathit{x'} \, , \, [ \, \mathit{x'} \, / \, \mathit{x} \, ] \, \ottnt{F} \, \vdash \, [ \, \ottkw{Suc} \, \mathit{x'} \, / \, \mathit{x} \, ] \, \ottnt{F} }%
}{
\Sigma \, ; \, \ottnt{H} \, \vdash \,  \forall  \mathit{x}  :  \ottkw{nat}  .    \ottkw{Terminates}\ \, \mathit{x}  \Rightarrow  \ottnt{F}   }{%
{\ottdrulename{Pv\_Ind}}{}%
}}

\newcommand{\ottdrulePvXXCompInd}[1]{\ottdrule[#1]{%
\ottpremise{ \Sigma \, , \, \mathit{f} \, : \,  \ottnt{A'}  \to  \ottnt{A}  \, ; \, \ottnt{H} \, , \,  \forall  \mathit{x}  :  \ottnt{A'}  .  [ \, \mathit{f} \, \mathit{x} \, / \, \mathit{z} \, ] \, \ottnt{F}  \, \vdash \,  \forall  \mathit{x}  :  \ottnt{A'}  .  [ \, \ottnt{t} \, / \, \mathit{z} \, ] \, \ottnt{F}  \ \ \ \ \   \Sigma  \vdash   \ottkw{rec} \  \mathit{f} ( \mathit{x} ) =  \ottnt{t}   :   \ottnt{A'}  \to  \ottnt{A}   }%
}{
\Sigma \, ; \, \ottnt{H} \, \vdash \,  \forall  \mathit{x}  :  \ottnt{A'}  .    \ottkw{Terminates}\ \, ( \,  \ottkw{rec} \  \mathit{f} ( \mathit{x} ) =  \ottnt{t}  \, ) \, \mathit{x}  \Rightarrow  [ \, ( \,  \ottkw{rec} \  \mathit{f} ( \mathit{x} ) =  \ottnt{t}  \, ) \, \mathit{x} \, / \, \mathit{z} \, ] \, \ottnt{F}   }{%
{\ottdrulename{Pv\_CompInd}}{}%
}}

\newcommand{\ottdrulePvXXTermZero}[1]{\ottdrule[#1]{%
}{
\Sigma \, ; \, \ottnt{H} \, \vdash \, \ottkw{Terminates}\ \, 0}{%
{\ottdrulename{Pv\_Term0}}{}%
}}

\newcommand{\ottdrulePvXXTermS}[1]{\ottdrule[#1]{%
\ottpremise{\Sigma \, ; \, \ottnt{H} \, \vdash \, \ottkw{Terminates}\ \, \ottnt{t}}%
}{
\Sigma \, ; \, \ottnt{H} \, \vdash \, \ottkw{Terminates}\ \, \ottkw{Suc} \, \ottnt{t}}{%
{\ottdrulename{Pv\_TermS}}{}%
}}

\newcommand{\ottdrulePvXXTermAbs}[1]{\ottdrule[#1]{%
}{
\Sigma \, ; \, \ottnt{H} \, \vdash \, \ottkw{Terminates}\ \, \lambda \, \mathit{x} \, . \, \ottnt{t}}{%
{\ottdrulename{Pv\_TermAbs}}{}%
}}

\newcommand{\ottdrulePvXXTermRec}[1]{\ottdrule[#1]{%
}{
\Sigma \, ; \, \ottnt{H} \, \vdash \, \ottkw{Terminates}\ \,  \ottkw{rec} \  \mathit{f} ( \mathit{x} ) =  \ottnt{t} }{%
{\ottdrulename{Pv\_TermRec}}{}%
}}

\newcommand{\ottdrulePvXXTermInv}[1]{\ottdrule[#1]{%
\ottpremise{\Sigma \, ; \, \ottnt{H} \, \vdash \, \ottkw{Terminates}\ \, {\cal C} \, [ \, \ottnt{t} \, ]}%
}{
\Sigma \, ; \, \ottnt{H} \, \vdash \, \ottkw{Terminates}\ \, \ottnt{t}}{%
{\ottdrulename{Pv\_TermInv}}{}%
}}

\newcommand{\ottdrulePvXXNotTermAbort}[1]{\ottdrule[#1]{%
\ottpremise{\Sigma \, ; \, \ottnt{H} \, \vdash \, \ottkw{Terminates}\ \, \ottkw{abort}}%
}{
\Sigma \, ; \, \ottnt{H} \, \vdash \, \ottnt{F}}{%
{\ottdrulename{Pv\_NotTermAbort}}{}%
}}

\newcommand{\ottdrulePvXXOpSem}[1]{\ottdrule[#1]{%
\ottpremise{ \ottnt{t}  \leadsto^*  \ottnt{t'} }%
}{
\Sigma \, ; \, \ottnt{H} \, \vdash \, \ottnt{t} \, = \, \ottnt{t'}}{%
{\ottdrulename{Pv\_OpSem}}{}%
}}

\newcommand{\ottdrulePvXXSubst}[1]{\ottdrule[#1]{%
\ottpremise{ \Sigma \, ; \, \ottnt{H} \, \vdash \, \ottnt{t} \, = \, \ottnt{t'} \ \ \ \ \  \Sigma \, ; \, \ottnt{H} \, \vdash \, [ \, \ottnt{t} \, / \, \mathit{x} \, ] \, \ottnt{F} }%
}{
\Sigma \, ; \, \ottnt{H} \, \vdash \, [ \, \ottnt{t'} \, / \, \mathit{x} \, ] \, \ottnt{F}}{%
{\ottdrulename{Pv\_Subst}}{}%
}}

\newcommand{\ottdefnpv}[1]{\begin{ottdefnblock}[#1]{$\Sigma \, ; \, \ottnt{H} \, \vdash \, \ottnt{F}$}{\ottcom{Proof rules for target logic}}
\ottusedrule{\ottdrulePvXXAssume{}}
\ottusedrule{\ottdrulePvXXAlli{}}
\ottusedrule{\ottdrulePvXXAlle{}}
\ottusedrule{\ottdrulePvXXImpi{}}
\ottusedrule{\ottdrulePvXXImpe{}}
\ottusedrule{\ottdrulePvXXAndi{}}
\ottusedrule{\ottdrulePvXXAndeOne{}}
\ottusedrule{\ottdrulePvXXAndeTwo{}}
\ottusedrule{\ottdrulePvXXTruei{}}
\ottusedrule{\ottdrulePvXXContra{}}
\ottusedrule{\ottdrulePvXXInd{}}
\ottusedrule{\ottdrulePvXXCompInd{}}
\ottusedrule{\ottdrulePvXXTermZero{}}
\ottusedrule{\ottdrulePvXXTermS{}}
\ottusedrule{\ottdrulePvXXTermAbs{}}
\ottusedrule{\ottdrulePvXXTermRec{}}
\ottusedrule{\ottdrulePvXXTermInv{}}
\ottusedrule{\ottdrulePvXXNotTermAbort{}}
\ottusedrule{\ottdrulePvXXOpSem{}}
\ottusedrule{\ottdrulePvXXSubst{}}
\end{ottdefnblock}}

\newcommand{\ottdefnsJtype}{
\ottdefnred{}
\ottdefnbeta{}
\ottdefnctxwf{}
\ottdefnsubeff{}
\ottdefnkassn{}
\ottdefntyassn{}
\ottdefnctxwfa{}
\ottdefnkassna{}
\ottdefntyassna{}
\ottdefnstassn{}
\ottdefnpv{}}

\newcommand{\ottdefnss}{
\ottdefnsJtype}

\newcommand{\ottall}{\ottmetavars\\[0pt]
\ottgrammar\\[5.0mm]

\ottdefnss
}

\renewcommand{\ottdrule}[4][]{{\displaystyle\frac{\begin{array}{l}#2\end{array}}{#3}\hspace{0.0cm}\ottdrulename{#4}}}

\maketitle

\begin{abstract}
This paper proposes a type-and-effect system called \Teqt, which
distinguishes terminating terms and total functions from possibly
diverging terms and partial functions, for a lambda calculus with
general recursion and equality types.  The central idea is to include
a primitive type-form ``Terminates t'', expressing that term t is
terminating; and then allow terms t to be coerced from possibly
diverging to total, using a proof of Terminates t.  We call such
coercions \emph{termination casts}, and show how to implement
terminating recursion using them.  For the meta-theory of the system,
we describe a translation from $\Teqt$ to a logical theory of
termination for general recursive, simply typed functions.  Every
typing judgment of $\Teqt$ is translated to a theorem expressing the
appropriate termination property of the computational part of the
$\Teqt$ term.
\end{abstract}

\section{Introduction}

Soundly combining general recursion and dependent types is a
significant current challenge in the design of dependently typed
programming languages.  The two main difficulties raised by this
combination are (1) type-equivalence checking with dependent types
usually depends on term reduction, which may fail to terminate in the
presence of general recursion; and (2) under the Curry-Howard
isomorphism, non-terminating recursions are interpreted as unsound
inductive proofs, and hence we lose soundness of the type system as a
logic.

Problem (1) can be addressed simply by bounding the number of steps of
reduction that can be performed in a single conversion.  This solution
may seem ad hoc, but it is less problematic if one works, as we do
here, with a primitive notion of propositional equality, and no
automatic conversion.  Explicit casts with equality proofs are used to
change the types of terms, and so with a bound on the number of
reduction steps allowed, one may simply chain together a sequence of
conversions to accommodate long-running terms in types.  There are
certainly some issues to be addressed in making such a solution
workable in practice, but it is not a fundamental problem.

Problem (2), on the other hand, cannot be so easily dealt with, since
we must truly know that a recursive function is total if we are to
view it soundly as an inductive proof.  One well-known approach to
this problem was proposed by Capretta~\cite{capretta05}: extend a
terminating type theory (that is, one for which we have a sound static
analysis for totality, which we use to require all functions to be
total) with general recursion via coinductive types.  Corecursion is
used to model general-recursive functions, without losing logical
soundness: productive corecursive functions correspond to sound
coinductive arguments.  The type constructor $(\cdot)^\nu$ for
possibly diverging computations, together with natural operations on
it, is shown to form a monad.


A separate problem related to (2) is extending the flexibility of
totality checking for total type theories.  It is well-known that
structural termination can become awkward for some functions like, for
example, natural-number division, where a recursive call must be made
on the result of another function call.  For this situation, methods like
type-based termination have been proposed: see Barthe et
al.~\cite{barthe+04} and several subsequent works by those authors;
also, Abel~\cite{abel06}.  The idea in type-based termination is,
roughly, to associate sizes with data, and track sizes statically
across function calls.  Recursive calls must be on data with smaller
size.  This method certainly increases the range of functions judged
total in their natural presentation.  No static termination analysis
will be complete, so there will always be programs that type-based
termination cannot judge terminating. When such analyses fail,
programmers must rewrite their code so that its termination behavior
is more apparent to the analysis. What is required is a flexible
method for such explicit termination arguments.

\paragraph{This paper's contribution}

This paper proposes a system called $\Teqt$ that can be seen as
building on both these lines of work.  We develop a type-and-effect
system where the effect distinguishes total from possibly partial
terms. The type assignment judgment \mbox{$ \Gamma  \vdash  \ottnt{t}  :  \ottnt{T} \  \theta $} includes a
{\em termination effect} $\theta$, which can be either $ \downarrow $ (called ``total''),
for terms that are known to terminate, or $ ? $ (called ``general''), for terms
whose termination behavior is unknown.

We can view this approach as building, at least in spirit, on
Capretta's approach with the partiality monad, thanks to the close
connection between monads and effects, as shown by Wadler and
Thiemann~\cite{wadler+03}.  Of course, there are
important differences between the monadic and effectful approaches,
most notably that effects are hard-wired into the language definition,
while monads are usually programmer-defined.  We adopt the effectful
approach here, since we are particularly focused on these two kinds of
computation, terminating and possibly partial, as fundamental.  We
thus deem them appropriate for hard-wiring into the language itself.
Exploring the tradeoffs more deeply between these two approaches must
remain to future work. 

   
%

Importantly, $\Teqt$ provides a flexible approach to termination
because the judgment of totality, \mbox{$ \Gamma  \vdash  \ottnt{t}  :  \ottnt{T} \   \downarrow  $}, is
internalized into the type system. The type $\ottkw{Terminates}\ \, \ottnt{t}$
expresses termination of term $\ottnt{t}$.  The effect of a term can thus be
changed from possibly partial to total by casting the term $\ottnt{t}$
with a proof of $\ottkw{Terminates}\ \, \ottnt{t}$.  These \emph{termination casts}
change the type checker's view of the termination behavior of a
term, much as a (sound) type cast changes its view of the type of the
term.  Termination casts are used with the terminating recursion
operator: the body of the putatively terminating recursive function is
type-checked under the additional explicit assumption that calls with
a structurally smaller argument are terminating.

By reifying this basic view of structural termination as an explicit
typing assumption, we follow the spirit of type-based termination: our
method eliminates the need for a separate structural check (proposed
as an important motivation for type-based
termination~\cite{barthe+04}), and gives the programmer even more
flexibility in the kind of functions s/he can write.  This is because
instead of relying on a static analysis to track sizes of datatypes,
our approach allows the user (or an automated reasoning system) to
perform arbitrarily complex reasoning to show termination of the
function.  This reasoning can be internal, using termination casts, or
completely external: one can write a general-recursive function that
the type checker can only judge to be possibly partial, and later
prove a theorem explicitly showing that the function is terminating.
Of course, one could also wish to support what we would see as a
hybrid approach, in the style of the \textsc{Program} tactic in Coq~\cite{sozeau06},
but this is outside the scope of the present paper.

\paragraph{Outline of the development} 

In Section~\ref{sec:lang}, we first present the syntax, reduction
rules and type assignment system for $\Teqt$.  Because type assignment
is not algorithmic for $\Teqt$, we also develop an annotated version
of $\Teqt$ suitable for implementation, where terms are annotated to
enable algorithmic type checking. We follow this
explanation with a number of examples of the use of termination casts,
in Section~\ref{sec:examples}.  Next, in Section~\ref{sec:consistency}
we develop our central meta-theoretic result, based on a translation
of $\Teqt$ typing judgments to judgments about termination of the term
in question, formulated in a first-order logical theory of
general-recursive functions (called $W'$).  This system is similar in
spirit to Feferman's theory $W$ (see Chapter 13 of~\cite{feferman98}),
although with significant syntactic differences, and support for
hypothetical reasoning about termination.  We show that $\Teqt$ is
sound with respect to this translation.  Also, we find that
constructive reasoning suffices for soundness of the translation, so
we take $W'$ to be intuitionistic (whereas an important characteristic
of $W$ is that its logic is classical).


\section{Definition of $\Teqt$}
\label{sec:lang}

The language \Teqt is a simple language with natural numbers and
dependently-typed recursive functions.  The syntax of types $\ottnt{T}$
and terms $\ottnt{t}$ appears in Figure~\ref{fig:syntax}. The variable
$\mathit{x}$ is bound in $\ottnt{t}$ in the term $\lambda \, \mathit{x} \, . \, \ottnt{t}$ and in $\ottnt{T'}$
in the type $ \Pi ^{ \theta }  \mathit{x} \!:\! \ottnt{T} .  \ottnt{T'} $.  As explained below, $\theta$
for $\Pi$-types represents the latent effect of the function's computation
(it does not describe the input argument).
The variables $\mathit{f}$ and $\mathit{x}$ are bound in $\ottnt{t}$ in the term
$ \ottkw{rec} \  \mathit{f} ( \mathit{x} ) =  \ottnt{t} $.  We use the notation $[ \, \ottnt{t'} \, / \, \mathit{x} \, ] \, \ottnt{T}$ and $[ \, \ottnt{t'} \, / \, \mathit{x} \, ] \, \ottnt{t}$ to denote the capture-avoiding substitution of
$\ottnt{t'}$ for $\mathit{x}$ in types and terms respectively.

We deliberately omit from \Teqt many important type-theoretic features
which we believe to be orthogonal to the central ideas explored here.
A full-fledged type theory based on these ideas would include
user-defined inductive types, type polymorphism, perhaps a universe
hierarchy, large eliminations, implicit products, and so forth.  Some
of these features, in particular large eliminations, raise serious
technical challenges for this approach (and many others). For
this paper we develop the core ideas needed for distinguishing
total and possibly partial computations with our effect system and
using termination casts to internalize termination, leaving other problems
to future work.

\begin{figure}
\[
\begin{array}{llcl}
\mathit{effects} & \theta,\rho & ::= &  \downarrow  \alt
 ?  \\
\mathit{types} & \ottnt{T} & ::= & \ottkw{nat} \alt  \Pi ^{ \theta }  \mathit{x} \!:\! \ottnt{T} .  \ottnt{T'}  \alt \ottnt{t} \, = \, \ottnt{t'} \alt \ottkw{Terminates}\ \, \ottnt{t} \\
\mathit{terms} & \ottnt{t} & ::= & \mathit{x} \alt \lambda \, \mathit{x} \, . \, \ottnt{t} \alt \ottnt{t} \, \ottnt{t'} \alt 0 \alt \ottkw{Suc} \, \ottnt{t} \\
&     &  \alt    &  \ottkw{rec} \  \mathit{f} ( \mathit{x} ) =  \ottnt{t}  \alt  \ottkw{case} \  \ottnt{t} \  \ottnt{t'} \  \ottnt{t''}  \\
&     &  \alt    & \ottkw{join} \alt \ottkw{terminates} \alt \ottkw{contra} \alt \ottkw{abort} \\
\mathit{values} & \ottnt{v} & ::=  & \mathit{x} \alt 0 \alt \ottkw{Suc} \, \ottnt{v} \alt \lambda \, \mathit{x} \, . \, \ottnt{t} 
 \alt  \ottkw{rec} \  \mathit{f} ( \mathit{x} ) =  \ottnt{t}  \\
& & \alt & \ottkw{join} \alt \ottkw{terminates} \alt \ottkw{contra} \\
\mathit{contexts} & {\cal C} & ::= & [] \alt \ottkw{Suc} \, {\cal C} \alt {\cal C} \, \ottnt{t} \alt \ottnt{v} \, {\cal C} \alt \ottkw{case} \, {\cal C} \, \ottnt{t} \, \ottnt{t} \\

\end{array}
\]
\caption{Syntax of $\Teqt$}
\label{fig:syntax}
\end{figure}

\subsection{Operational semantics}

Reduction for \Teqt is defined as a call-by-value small-step
operational semantics. Figure~\ref{fig:syntax} presents the syntax of
values and evaluation contexts and Figure~\ref{fig:op-sem} contains
the two judgments that make up this semantics. Values in \Teqt include
variables, natural numbers, functions and primitive proof terms for
the internalized judgments of equality and termination.

We define the reduction rules with two relations: the primitive
$\beta$ rules, written $\ottnt{t} \, \leadsto_\beta \, \ottnt{t'}$ describe reduction when a value
is in an active position. This relation is used by the main reduction
relation $\ottnt{t} \, \leadsto \, \ottnt{t'}$, which lifts beta reduction through evaluation
contexts ${\cal C}$ and terminates computation for $\ottkw{abort}$,
representing finite failure.  Other proof forms, including
$\ottkw{contra}$, are considered values.  We cannot, in fact, obtain a
contradiction in the empty context (assuming our theory $W'$ is
consistent), but at this point in the development that cannot be
shown.  

\begin{figure}
\begin{minipage}[t]{0.6\linewidth}
\centering
\ottdefnbeta{}
\end{minipage}
\begin{minipage}[t]{0.4\linewidth}
\centering
\ottdefnred{}
\end{minipage}
\caption{Call-by-value small-step operational semantics}
\label{fig:op-sem}
\end{figure}

\subsection{Type assignment} Figure~\ref{fig:tpassn} defines the {\em
type-assignment} system.  The judgment $ \Gamma  \vdash  \ottnt{t}  :  \ottnt{T} \  \theta $ states
that the term $\ottnt{t}$ can be assigned type $\ottnt{T}$ in the context
$\Gamma$ with effect $\theta$. (The other two judgments, $\Gamma \, \vdash \, \ottkw{Ok}$ and $ \Gamma  \vdash  \ottnt{T} $, are used by this one to check that
contexts and types are well formed.)  We define the system such that
$\theta$ is an approximation of the termination behavior of the
system. If we can derive a judgment $ \Gamma  \vdash  \ottnt{t}  :  \ottnt{T} \   \downarrow  $, then this
means that for any assignment of values to the variables in $\Gamma$,
reduction of $\ottnt{t}$ must terminate.  (If the context is inconsistent,
$\ottnt{t}$ might not terminate even if the type system judges it to do
so, since an inconsistent context can make unsatisfiable assertions
about termination, which may pollute the type system's judgments.) In
contrast, the judgment $ \Gamma  \vdash  \ottnt{t}  :  \ottnt{T} \   ?  $ places no restrictions
on the termination behavior of $\ottnt{t}$. We
view $\theta$ is as a {\em capability} on termination
behavior~\cite{crary+99}. A term with capability $ ? $ is
allowed to diverge, but terms with capability $ \downarrow $ cannot.  As
a result, any term that typechecks with $ \downarrow $ will also
typecheck with $ ? $.  Thus $ ? $ is more permissive
than $ \downarrow $, and we order them as $ \downarrow  \, \leq \,  ? $.

\begin{figure}
\[
\begin{array}{ll}
\fbox{$ \Gamma  \vdash  \ottnt{T} $} \\ \\
\ottdrulekXXNat{} & \ottdrulekXXPi{} \\  \\
\ottdrulekXXEq{}  & \ottdrulekXXTerm{} \\ \\
\fbox{$\Gamma \, \vdash \, \ottkw{Ok}$ } \\ \\
\ottdruleOkXXempty{} & \ottdruleOkXXcons{} \\ \\ 
\fbox{$ \Gamma  \vdash  \ottnt{t}  :  \ottnt{T} \  \theta $} \\ \\
\ottdruleTXXJoin{} & \ottdruleTXXConv{} \\ \\
\ottdruleTXXReify{} & \ottdruleTXXReflect{} \\ \\
\ottdruleTXXCtxTerm{} & \ottdruleTXXVar{}  \\ \\
\ottdruleTXXAbs{} & \ottdruleTXXApp{} \\ \\
\ottdruleTXXZero{} & \ottdruleTXXSuc{} \\ \\
\ottdruleTXXContra{} & \ottdruleTXXAbort{}  \\ \\
\ottdruleTXXRec{} & \ottdruleTXXCase{} \\
\end{array}
\]
\[
\begin{array}{l}
\ottdruleTXXRecNat{} \\ 
\end{array}
\]
\caption{Type assignment system}
\label{fig:tpassn}
\end{figure}

Such reasoning is reflected in the type system. $\Teqt$ has a
call-by-value operational semantics, so variables stand for
values. Therefore, a variable is known to terminate, so we can type
variables with any effect in rule \ottdrulename{T\_Var}. This pattern
occurs often; all terms that are known to terminate have unconstrained
effects in the conclusion of their typing rules. In this way, we build
subeffecting into the type system and do not need an additional rule
to coerce total terms to general ones. Because of this subeffecting,
when a premise of a rule uses the general effect, such as
\ottdrulename{K\_Eq}, it places no restriction on the term.

As is standard in type-and-effect systems, function types are
annotated with a {\em latent effect}. This effect records the
termination effect for the body of the function, in rule
\ottdrulename{T\_Abs}.  Likewise, in an application (rule
\ottdrulename{T\_App}), the latent effect of the function must be
equal or less than the current termination effect. Note that, although
the system supports subeffecting, it does not support subtyping. In an
application, the type of the argument must exactly match that expected
by the function. Although there is a natural extension of subeffecting
to subtyping, for simplicity we have not included it in this system.

\Teqt types include two propositions. The type $\ottnt{t} \, = \, \ottnt{t'}$ states
that two terms are equal and the type $\ottkw{Terminates}\ \, \ottnt{t}$ declares that
term $\ottnt{t}$ is terminating.  The introduction form for the equality
proposition (rule \ottdrulename{T\_Join}) requires both terms to be
well typed and evaluate to a common reduct. For flexibility, these
terms need not be judged terminating nor have the same type. The
elimination form (\ottdrulename{T\_Conv}) uses a total proof of
equality to convert between equivalent types.  Likewise, the
introduction form for the $\ottkw{Terminates}\ \, \ottnt{t}$ proposition
(\ottdrulename{T\_Reify}) requires showing that the term terminates.
Analogously, the elimination form (\ottdrulename{T\_Reflect}) uses a
total proof of termination to change the effect of $\ottnt{t}$.  \Teqt
also internalizes an admissible property of the judgment with the
empty context---if a term terminates, then the subterm in the active
position of the term terminates (\ottdrulename{T\_CtxTerm}).  This
property does not (appear to) follow constructively from the others.

Recursive functions can be typed with either general or total latent
effects. In the latter case, the \ottdrulename{T\_RecNat} rule
introduces a new hypothesis into the context that may be used to
show that the body of the function is total. The assumption $\mathit{p} :
 \Pi ^{  \downarrow  }  \mathit{x_{{\mathrm{1}}}} \!:\! \ottkw{nat} .   \Pi ^{  \downarrow  }  \mathit{p'} \!:\! \mathit{x} \, = \, \ottkw{Suc} \, \mathit{x_{{\mathrm{1}}}} .  \ottkw{Terminates}\ \, ( \, \mathit{f} \, \mathit{x_{{\mathrm{1}}}} \, )  $ is
an assertion that for any number $\mathit{x_{{\mathrm{1}}}}$ that is one less than
$\mathit{x}$, the recursive call $( \, \mathit{f} \, \mathit{x_{{\mathrm{1}}}} \, )$ terminates. Even though the
type of $\mathit{f}$ has a $ ? $ latent effect, recursive calls
on the immediate predecessor can be cast to be total using this assumption.

The rule \ottdrulename{T\_RecNat} includes a restriction that 
$\mathit{p} \, \not\in \, \ottkw{fv} \, \ottnt{t}$. This means that the only places that $\mathit{p}$ can occur
in a typing derivation is in the proof-premises of
\ottdrulename{T\_Conv}, \ottdrulename{T\_Reflect}, and
\ottdrulename{T\_Contra}. The advantage of setting up the system this way is that we can
define the operational semantics without any reference to proofs: the rule \ottdrulename{Beta\_AppRec}
does not have to specify a proof term to substitute for free occurrences of $\mathit{p}$ in $\ottnt{t}$.
In other words the \ottdrulename{T\_RecNat} rule bakes in a form of \emph{proof erasure}~\cite{mishra-linger+08,barras+08,miquel01}.

We may worry that this restriction limits the expressiveness
of the language because the variable $\mathit{p}$ can not be used in every
context. However, that is not the case as our system satisfies a form
of \emph{proof irrelevance}. No matter what proof we have of
termination, we can always use the rules \ottdrulename{T\_Reify} and
\ottdrulename{T\_Reflect} to replace it by the (computationally)
uninformative proof $\ottkw{terminates}$. We give an example of this
behavior in the next section. Thus, we do not lose anything by
making the proof variable $\mathit{p}$ second-class, since we can always
replace it with a proof that does not mention $\mathit{p}$.
(Likewise, equality proofs are irrelevant, as we can use
\ottdrulename{T\_Join} followed by \ottdrulename{T\_Conv} to show that 
$ \Gamma  \vdash  \mathit{u}  :  \ottnt{t} \, = \, \ottnt{t'} \   \downarrow  $ implies $ \Gamma  \vdash  \ottkw{join}  :  \ottnt{t} \, = \, \ottnt{t'} \   \downarrow  $.)

\subsection{Annotated language}
\label{sec:anno}
 
The previous two subsections provide a complete specification of the
\Teqt language. However, in \Teqt, type inference is not
algorithmic. Given a context $\Gamma$, a term $\ottnt{t}$ and effect
$\theta$, it is not clear how to determine if there is some $\ottnt{T}$
such that $ \Gamma  \vdash  \ottnt{t}  :  \ottnt{T} \  \theta $ holds. The terms do not contain
enough information to indicate how to construct a typing derivation.

Fortunately, it is straightforward to produce an annotated version of
$\Teqt$ where the type checking algorithm is fully determined. Below
we give the syntax of the annotated terms. The full typing rules for
the annotated system appear in Figure~\ref{fig:tpannot}.  The
judgment form is $ \Gamma  \Vdash  \ottnt{a}  :  \ottnt{S} \  \theta $, where algorithmically,
$\Gamma$, $\ottnt{a}$, and $\theta$ are inputs to the type checker
and type $\ottnt{S}$ is the output.

\begin{figure}
\[
\begin{array}{lrcl}
\mathit{annot.\ types} &S &::=&  \ottkw{nat} \alt  \Pi ^{ \theta }  x \!:\! \ottnt{S} .  \ottnt{S'}  \alt \ottnt{a} \, = \, \ottnt{a'} \alt \ottkw{Terminates}\ \, \ottnt{a} \\
\mathit{annot.\ terms} &a &::=&  x \alt \ottnt{a} \, \ottnt{a'} \alt  \lambda^{\textcolor{black}{ \theta } }  x  \textcolor{black}{\!:\!  \ottnt{S} }. \ottnt{a}  \alt 0 \alt \ottkw{Suc} \, \ottnt{a} \\
&  &\alt&  \ottkw{rec} _{\textcolor{black}{ \ottkw{nat} } }\  f ( x \  p ) \textcolor{black}{ :  \ottnt{S}  } =  \ottnt{a}  \alt  \ottkw{rec} \  f ( x  \textcolor{black}{\!:\! \ottnt{S} }) \textcolor{black}{: \ottnt{S'} } =  \ottnt{a}  \alt 
          \ottkw{case} \ \textcolor{black}{ x  .  \ottnt{S} }\  \ottnt{a} \  \ottnt{a'} \  \ottnt{a''}  \\
&  &\alt&  \ottkw{join} \ \textcolor{black}{ \ottnt{a} \  \ottnt{a'} }  \alt  \textcolor{black}{ \ottkw{conv} \  x . \ottnt{S} }\  \ottnt{a'} \ \textcolor{black}{ \ottnt{a} }  \alt  \ottkw{terminates} \ \textcolor{black}{ \ottnt{a} }  \alt  \textcolor{black}{ \ottkw{reflect} }\  \ottnt{a} \ \textcolor{black}{ \ottnt{a'} }  \\ 
&  &\alt&  \textcolor{black}{ \ottkw{inv} }\  \ottnt{a} \ \textcolor{black}{ \ottnt{a'} }  \alt  \ottkw{contra} \ \textcolor{black}{  \ottnt{S} \  \ottnt{a}  }  \alt  \ottkw{abort} \ \textcolor{black}{  \ottnt{S} }  \\
\end{array}
\]
\caption{Syntax of annotated $\Teqt$}
\label{fig:annot-syntax}
\end{figure}

\begin{figure}
\[
\begin{array}[t]{lcl}
\mathit{Types}\\
\ottkw{\mbox{$\mid$}} \, \ottkw{nat} \, \ottkw{\mbox{$\mid$}} &=& \ottkw{nat} \\
\ottkw{\mbox{$\mid$}} \,  \Pi ^{ \theta }  x \!:\! \ottnt{S} .  \ottnt{S'}  \, \ottkw{\mbox{$\mid$}} &=&  \Pi ^{ \theta }  \mathit{x} \!:\! \ottkw{\mbox{$\mid$}} \, \ottnt{S} \, \ottkw{\mbox{$\mid$}} .  \ottkw{\mbox{$\mid$}} \, \ottnt{S'} \, \ottkw{\mbox{$\mid$}}  \\
\ottkw{\mbox{$\mid$}} \, \ottnt{a} \, = \, \ottnt{a'} \, \ottkw{\mbox{$\mid$}} &=& \ottkw{\mbox{$\mid$}} \, \ottnt{a} \, \ottkw{\mbox{$\mid$}} \, = \, \ottkw{\mbox{$\mid$}} \, \ottnt{a'} \, \ottkw{\mbox{$\mid$}} \\
\ottkw{\mbox{$\mid$}} \, \ottkw{Terminates}\ \, \ottnt{a} \, \ottkw{\mbox{$\mid$}} &=& \ottkw{Terminates}\ \, \ottkw{\mbox{$\mid$}} \, \ottnt{a} \, \ottkw{\mbox{$\mid$}} \\
\end{array}
\]
\[
\begin{array}{cc}
\begin{array}[t]{lcl}
\mathit{Terms}\\
\ottkw{\mbox{$\mid$}} \, x \, \ottkw{\mbox{$\mid$}} &=& \mathit{x} \\
\ottkw{\mbox{$\mid$}} \, \ottnt{a} \, \ottnt{a'} \, \ottkw{\mbox{$\mid$}} &=& \ottkw{\mbox{$\mid$}} \, \ottnt{a} \, \ottkw{\mbox{$\mid$}} \, \ottkw{\mbox{$\mid$}} \, \ottnt{a'} \, \ottkw{\mbox{$\mid$}} \\
\ottkw{\mbox{$\mid$}} \,  \lambda^{\textcolor{black}{ \theta } }  x  \textcolor{black}{\!:\!  \ottnt{S} }. \ottnt{a}  \, \ottkw{\mbox{$\mid$}} &=& \lambda \, \mathit{x} \, . \, \ottkw{\mbox{$\mid$}} \, \ottnt{a} \, \ottkw{\mbox{$\mid$}} \\
\ottkw{\mbox{$\mid$}} \, 0 \, \ottkw{\mbox{$\mid$}} &=& 0 \\ 
\ottkw{\mbox{$\mid$}} \, \ottkw{Suc} \, \ottnt{a} \, \ottkw{\mbox{$\mid$}} &=& \ottkw{Suc} \, \ottkw{\mbox{$\mid$}} \, \ottnt{a} \, \ottkw{\mbox{$\mid$}} \\
\ottkw{\mbox{$\mid$}} \,  \ottkw{case} \ \textcolor{black}{ x  .  \ottnt{S} }\  \ottnt{a} \  \ottnt{a'} \  \ottnt{a''}  \, \ottkw{\mbox{$\mid$}} &=&  \ottkw{case} \  \ottkw{\mbox{$\mid$}} \, \ottnt{a} \, \ottkw{\mbox{$\mid$}} \  \ottkw{\mbox{$\mid$}} \, \ottnt{a'} \, \ottkw{\mbox{$\mid$}} \  \ottkw{\mbox{$\mid$}} \, \ottnt{a''} \, \ottkw{\mbox{$\mid$}}  \\
\ottkw{\mbox{$\mid$}} \,  \ottkw{rec} _{\textcolor{black}{ \ottkw{nat} } }\  f ( x \  p ) \textcolor{black}{ :  \ottnt{S}  } =  \ottnt{a}  \, \ottkw{\mbox{$\mid$}} &=&  \ottkw{rec} \  \mathit{f} ( \mathit{x} ) =  \ottkw{\mbox{$\mid$}} \, \ottnt{a} \, \ottkw{\mbox{$\mid$}}  \\
\ottkw{\mbox{$\mid$}} \,  \ottkw{rec} \  f ( x  \textcolor{black}{\!:\! \ottnt{S} }) \textcolor{black}{: \ottnt{S'} } =  \ottnt{a}  \, \ottkw{\mbox{$\mid$}} &=&  \ottkw{rec} \  \mathit{f} ( \mathit{x} ) =  \ottkw{\mbox{$\mid$}} \, \ottnt{a} \, \ottkw{\mbox{$\mid$}}  \\
\end{array}
&
\begin{array}[t]{lcl}
\\
\ottkw{\mbox{$\mid$}} \,  \ottkw{join} \ \textcolor{black}{ \ottnt{a} \  \ottnt{a'} }  \, \ottkw{\mbox{$\mid$}} &=& \ottkw{join} \\
\ottkw{\mbox{$\mid$}} \,  \ottkw{terminates} \ \textcolor{black}{ \ottnt{a} }  \, \ottkw{\mbox{$\mid$}} &=& \ottkw{terminates} \\
\ottkw{\mbox{$\mid$}} \,  \ottkw{contra} \ \textcolor{black}{  \ottnt{S} \  \ottnt{a}  }  \, \ottkw{\mbox{$\mid$}} &=& \ottkw{contra} \\
\ottkw{\mbox{$\mid$}} \,  \ottkw{abort} \ \textcolor{black}{  \ottnt{S} }  \, \ottkw{\mbox{$\mid$}} &=& \ottkw{abort} \\
\ottkw{\mbox{$\mid$}} \,  \textcolor{black}{ \ottkw{conv} \  x . \ottnt{S} }\  \ottnt{a} \ \textcolor{black}{ \ottnt{a'} }  \, \ottkw{\mbox{$\mid$}} &=& \ottkw{\mbox{$\mid$}} \, \ottnt{a} \, \ottkw{\mbox{$\mid$}} \\
\ottkw{\mbox{$\mid$}} \,  \textcolor{black}{ \ottkw{reflect} }\  \ottnt{a} \ \textcolor{black}{ \ottnt{a'} }  \, \ottkw{\mbox{$\mid$}} &=& \ottkw{\mbox{$\mid$}} \, \ottnt{a} \, \ottkw{\mbox{$\mid$}} \\
\ottkw{\mbox{$\mid$}} \,  \textcolor{black}{ \ottkw{inv} }\  \ottnt{a} \ \textcolor{black}{ \ottnt{a'} }  \, \ottkw{\mbox{$\mid$}} &=& \ottkw{\mbox{$\mid$}} \, \ottnt{a} \, \ottkw{\mbox{$\mid$}} \\
\end{array}
\end{array}
\]
\caption{Annotation erasure}
\label{fig:erasure}
\end{figure}

\begin{figure}
\[
\begin{array}{ll}
\fbox{$ \Gamma  \Vdash  \ottnt{S} $} \\ \\
\ottdruleSXXNat{} & \ottdruleSXXPi{} \\  \\
\ottdruleSXXEq{}  & \ottdruleSXXTerm{} \\ \\
\fbox{$ \Gamma  \Vdash  \ottkw{Ok} $ } \\ 
\ottdruleOkaXXempty{} & \ottdruleOkaXXcons{} \\ \\ 
\fbox{$ \Gamma  \Vdash  \ottnt{a}  :  \ottnt{S} \  \theta $} \\ \\
\ottdruleAXXJoin{} & \ottdruleAXXConv{} \\ \\
\ottdruleAXXReify{} & \hspace{-0.2cm}\ottdruleAXXReflect{} \\ \\
\ottdruleAXXCtxTerm{} & \ottdruleAXXVar{}  \\ \\
\ottdruleAXXAbs{} & \hspace{-0.3cm}\ottdruleAXXApp{} \\ \\
\ottdruleAXXZero{} & \ottdruleAXXSuc{} \\ \\
\ottdruleAXXContra{} & \ottdruleAXXAbort{}  \\ \\
\ottdruleAXXRec{} & \ottdruleAXXCase{} \\
\end{array}
\]
\[
\begin{array}{l}
\hspace{-0.2cm}\ottdruleAXXRecNat{} \\ 
\end{array}
\]
\caption{Annotated type checking system}
\label{fig:tpannot}
\end{figure}

Most annotated term forms have direct correspondence to the
unannotated terms. Figure~\ref{fig:erasure} defines the operation
$|\cdot|$ that erases annotations. Notably, there are two different forms of
recursion, based on which typing rule should be used. Furthermore, the
syntax includes terms ($ \textcolor{black}{ \ottkw{conv} \  x . \ottnt{S} }\  \ottnt{a'} \ \textcolor{black}{ \ottnt{a} } $, $ \textcolor{black}{ \ottkw{inv} }\  \ottnt{a} \ \textcolor{black}{ \ottnt{a'} } $, and
$ \textcolor{black}{ \ottkw{reflect} }\  \ottnt{a} \ \textcolor{black}{ \ottnt{a'} } $) that mark where type conversions, termination
inversions and termination casts should occur---these are implicit in
the unannotated system.

The annotated system uses types $\ottnt{S}$ that are exactly like types
$\ottnt{T}$ except that they contain annotated terms.  However, because
there is no operational semantics defined for annotated terms, the
join rule (shown below) first erases the annotations before
determining if there is some common reduct. Likewise, the
inversion rule uses erasure to find the evaluation context.

Simple comparison of the typing rules of the two systems in a
straightforward inductive proof shows that the annotated system is
sound and complete with respect to the implicit system.

\begin{proposition}[Soundness of annotated system]
If $ \Gamma  \Vdash  \ottnt{a}  :  \ottnt{S} \  \theta $ then $ \Gamma  \vdash  \ottkw{\mbox{$\mid$}} \, \ottnt{a} \, \ottkw{\mbox{$\mid$}}  :  \ottkw{\mbox{$\mid$}} \, \ottnt{S} \, \ottkw{\mbox{$\mid$}} \  \theta $.
\end{proposition}
\begin{proposition}[Completeness of annotated system]
If $ \Gamma  \vdash  \ottnt{t}  :  \ottnt{T} \  \theta $ then there exists an $\ottnt{a}$ and $\ottnt{S}$, such 
that $\ottkw{\mbox{$\mid$}} \, \ottnt{a} \, \ottkw{\mbox{$\mid$}}=\ottnt{t}$ and $\ottkw{\mbox{$\mid$}} \, \ottnt{S} \, \ottkw{\mbox{$\mid$}}=\ottnt{T}$ and $ \Gamma  \Vdash  \ottnt{a}  :  \ottnt{S} \  \theta $.
\end{proposition}

Note that although type inference is syntax-directed, it is only
decidable in the annotated system if there is some cut-off in
normalization in the join rule.  Even if we were to require $\ottnt{a}$
and $\ottnt{a'}$ to have the total effect in this rule, this restriction
would not ensure decidability. An inconsistent context could type a
looping term with a total effect.  It would be reasonable to make the
cutoff part of the annotated $\ottkw{join}$-term itself, although here
we use a global cut-off. Note that imposing a cutoff in the join rule
in the annotated system does not jeopardize completeness as a single
join in the implicit system can be translated to several joins in
the annotated system.

Finally, we are not considering the problem of annotation inference
for this system.  This is an important problem to ease the burden of
programming with termination casts.  We conjecture that in many simple
cases like structural decrease of a single parameter to the function,
the appropriate termination casts can be added completely
automatically.  But working this process out is beyond the scope of this
paper.

\section{Examples} \label{sec:examples} 

\paragraph{Natural number addition: internal verification}
Our first example shows how simple structurally recursive functions
can be shown terminating at their definition time using the
\ottdrulename{T\_RecNat} rule. 
We define natural number addition with the following term, showing 
first its implicit then annotated versions: 
\newcommand\erase[1]{\textcolor{black}{#1}}

\[
\begin{array}{ll}
\mathit{implicit }\  \mathit{plus}  &\defeq
\lambda \, \mathit{x_{{\mathrm{2}}}} \, . \,  \ottkw{rec} \  \mathit{f} ( \mathit{x_{{\mathrm{1}}}} ) =  ( \,  \ottkw{case} \  \mathit{x_{{\mathrm{1}}}} \  ( \, \lambda \, \mathit{q} \, . \, \mathit{x_{{\mathrm{2}}}} \, ) \  ( \, \lambda \, \mathit{x'} \, . \, \lambda \, \mathit{q} \, . \, \ottkw{Suc} \, ( \, \mathit{f} \, \mathit{x'} \, ) \, )  \, ) \, \ottkw{join} \\
\mathit{annotated }\  \mathit{plus}  &\defeq
\!\!\begin{array}[t]{l}
 \lambda ^{\erase{ \downarrow }} x_{{\mathrm{2}}}\erase{\!:\!\ottkw{nat}}. \   \ottkw{rec}_{\ottkw{nat} } \ f\ ( x_{{\mathrm{1}}}\ p ) \erase{: \ottkw{nat} }  =  \\
\quad  (  \ottkw{case} \  \erase{x. ( \,  \Pi ^{  \downarrow  }  q \!:\! x_{{\mathrm{1}}} \, = \, x .  \ottkw{nat}  \, )}
\  x_{{\mathrm{1}}} \\
\quad\quad     ( \,  \lambda^{\textcolor{black}{  \downarrow  } }  q  \textcolor{black}{\!:\!  x_{{\mathrm{1}}} \, = \, 0 }. x_{{\mathrm{2}}}  \, ) \\
\quad\quad     ( \lambda ^{\erase{ \downarrow }}x'\erase{\!:\!\ottkw{nat}} .  \lambda ^{\erase{ \downarrow }}q\erase{\!:\!x_{{\mathrm{1}}} \, = \, \ottkw{Suc} \, x'}. \  \ottkw{Suc} \, ( \,  \textcolor{black}{ \ottkw{reflect} }\  ( \, f \, x' \, ) \ \textcolor{black}{ ( \, p \, x' \, q \, ) }  \, ))) \\
\quad ( \,  \ottkw{join} \ \textcolor{black}{ x_{{\mathrm{1}}} \  x_{{\mathrm{1}}} }  \, )
\end{array}
\end{array}
\]
In this example, we must abstract over equality types that are then
applied to $\ottkw{join}$. This standard trick, used frequently in
\textsc{Coq} and similar dependent type theories, introduces different
assumptions of equalities into the context, depending on the case
branch.  As remarked above, we have deliberately omitted from \Teqt a
number of features that would improve some of these examples, notably
implicit products (as proposed by Miquel~\cite{miquel01}) for equality
proofs in case-terms.

The typing rules verify that plus is a total operation. For example, 
in the annotated system we can show:
\[
  \cdot   \Vdash   \mathit{plus}   :   \Pi ^{  \downarrow  }  x_{{\mathrm{1}}} \!:\! \ottkw{nat} .   \Pi ^{  \downarrow  }  x_{{\mathrm{2}}} \!:\! \ottkw{nat} .  \ottkw{nat}   \   \downarrow  
\]
To see why this is so, consider the context that we use to type check 
the body of the recursive function:
\[ 
\begin{array}{l}
\Gamma \defeq 
x_{{\mathrm{1}}} \, : \, \ottkw{nat} \, , \, x_{{\mathrm{2}}} \, : \, \ottkw{nat} \, , \, f \, : \,  \Pi ^{  ?  }  x_{{\mathrm{1}}} \!:\! \ottkw{nat} .  \ottkw{nat}  \, , \, p \, : \,  \Pi ^{  \downarrow  }  x' \!:\! \ottkw{nat} .   \Pi ^{  \downarrow  }  q \!:\! x_{{\mathrm{1}}} \, = \, \ottkw{Suc} \, x' .  \ottkw{Terminates}\ \, ( \, f \, x' \, )   \, , \,  \cdot 
\end{array}
\]
In this context, we would like to show that the case expression has type 
$ ( \,  \Pi ^{  \downarrow  }  q \!:\! x_{{\mathrm{1}}} \, = \, x_{{\mathrm{1}}} .  \ottkw{nat}  \, ) $. 
Note that the abstraction of $q$ must be $ \downarrow $ so that when we apply
the case expression to $\ottkw{join}$ the entire expression will have the $ \downarrow $ effect.
In the zero case, we use rules \ottdrulename{TA\_Abs} and
\ottdrulename{TA\_Var} to show that the abstraction has the desired
total function type.

In the successor case, we use a termination cast to show that the
recursive call is total. Without this cast, we would be
unable to use the latent effect $ \downarrow $ in the abstraction of
$q$. Using the rules for variables and application we can show that 
the recursive call has a general effect, but by itself, this will not 
let us define a total function.
\[
  \Gamma \, , \, x' \, : \, \ottkw{nat} \, , \, q \, : \, x_{{\mathrm{1}}} \, = \, \ottkw{Suc} \, x'  \Vdash  f \, x'  :  \ottkw{nat} \   ?   
\] 
However, given the extra argument from recursive function, we can 
produce a proof that the recursive call terminates.
\[ 
  \Gamma \, , \, x' \, : \, \ottkw{nat} \, , \, q \, : \, x_{{\mathrm{1}}} \, = \, \ottkw{Suc} \, x'  \Vdash  p \, x' \, q  :  \ottkw{Terminates}\ \, ( \, f \, x' \, ) \   \downarrow   
\]
From these two, we can use a termination cast to change the 
effect of the recursive call.
\[
  \Gamma \, , \, x' \, : \, \ottkw{nat} \, , \, q \, : \, x_{{\mathrm{1}}} \, = \, \ottkw{Suc} \, x'  \Vdash   \textcolor{black}{ \ottkw{reflect} }\  ( \, f \, x' \, ) \ \textcolor{black}{ ( \, p \, x' \, q \, ) }   :  \ottkw{nat} \   \downarrow  
\]
Finally, we can use the rules for successor and abstraction to
conclude that the successor case has the desired type.

\paragraph{Natural number addition: external verification}
An advantage of this system is that we do not need to prove that 
plus is total when we define it. We could also define plus using 
general recursion:
\[ 
  \mathit{plus}  \defeq \lambda \, \mathit{x_{{\mathrm{2}}}} \, . \,  \ottkw{rec} \  \mathit{f} ( \mathit{x_{{\mathrm{1}}}} ) =   \ottkw{case} \  \mathit{x_{{\mathrm{1}}}} \  \mathit{x_{{\mathrm{2}}}} \  ( \, \lambda \, \mathit{z} \, . \, \ottkw{Suc} \, ( \, \mathit{f} \, \mathit{z} \, ) \, )  
\]
But note, the best typing derivation will assign a $ ? $
latent effect to this function.  (For brevity, this and further
examples will be presented in the implicit language.)
\[
  \cdot   \vdash   \mathit{plus}   :   \Pi ^{  \downarrow  }  \mathit{x_{{\mathrm{2}}}} \!:\! \ottkw{nat} .   \Pi ^{  ?  }  \mathit{x_{{\mathrm{1}}}} \!:\! \ottkw{nat} .  \ottkw{nat}   \   \downarrow  
\]
However, all is not lost.  We can still prove the following theorem
and use it in a termination cast to show that a particular
application of $ \mathit{plus} $ terminates. The proof term (below) uses recursion to construct a total witness for this 
theorem.
\[
\begin{array}{lcl}
\mathit{plustotal}& : & 
 \Pi ^{  \downarrow  }  \mathit{x_{{\mathrm{2}}}} \!:\! \ottkw{nat} .   \Pi ^{  \downarrow  }  \mathit{x_{{\mathrm{1}}}} \!:\! \ottkw{nat} .  \ottkw{Terminates}\ \, ( \,  \mathit{plus}  \, \mathit{x_{{\mathrm{2}}}} \, \mathit{x_{{\mathrm{1}}}} \, )   \\
\mathit{plustotal}&\defeq & \lambda \, \mathit{x_{{\mathrm{2}}}} \, . \, ( \,  \ottkw{rec} \  \mathit{f} ( \mathit{x_{{\mathrm{1}}}} ) =  ( \,  \ottkw{case} \  \mathit{x_{{\mathrm{1}}}} \  ( \, \lambda \, \mathit{q} \, . \, \ottkw{terminates} \, ) \  ( \, \lambda \, \mathit{z} \, . \, \lambda \, \mathit{q} \, . \, \ottkw{terminates} \, )  \, ) \, \ottkw{join}  \, )
\end{array}
\]
To understand this proof term, we look at the typing derivation in
each branch of the case term. Let $\Gamma$ be the context  
that rule \ottdrulename{T\_RecNat} uses to check the body of the 
recursive definition, shown below.
\[ \Gamma \defeq
\begin{array}[t]{rcl}
\mathit{x_{{\mathrm{2}}}} &:& \ottkw{nat}, \\
 \mathit{x_{{\mathrm{1}}}} &:& \ottkw{nat}, \\
 \mathit{f} &:&  \Pi ^{  ?  }  \mathit{z} \!:\! \ottkw{nat} .  \ottkw{Terminates}\ \, ( \,  \mathit{plus}  \, \mathit{x_{{\mathrm{2}}}} \, \mathit{z} \, ) ,\\
 \mathit{p} &:&  \Pi ^{  \downarrow  }  \mathit{z} \!:\! \ottkw{nat} .   \Pi ^{  \downarrow  }  \mathit{q} \!:\! \mathit{x_{{\mathrm{1}}}} \, = \, \ottkw{Suc} \, \mathit{z} .  \ottkw{Terminates}\ \, ( \, \mathit{f} \, \mathit{z} \, )  \\
\end{array}
\]
 Then in the zero case, because $ \mathit{plus}  \, \mathit{x_{{\mathrm{2}}}} \, 0$
evaluates to $\mathit{x_{{\mathrm{2}}}}$ and variables terminate, we can use rule
\ottdrulename{T\_Conv} to show that case total.
\[
\cfrac{
\cfrac{  \Gamma \, , \, \mathit{q} \, : \, \mathit{x_{{\mathrm{1}}}} \, = \, 0  \vdash  \mathit{x_{{\mathrm{2}}}}  :  \ottkw{nat} \   \downarrow   }{
 \Gamma \, , \, \mathit{q} \, : \, \mathit{x_{{\mathrm{1}}}} \, = \, 0  \vdash  \ottkw{terminates}  :  \ottkw{Terminates}\ \, \mathit{x_{{\mathrm{2}}}} \   \downarrow  } \qquad 
\cfrac{\vdots}{ \Gamma  \vdash  \ottkw{join}  :   \mathit{plus}  \, \mathit{x_{{\mathrm{2}}}} \, 0 \, = \, \mathit{x_{{\mathrm{2}}}} \   \downarrow   }
}{\cfrac{
 \Gamma \, , \, \mathit{q} \, : \, \mathit{x_{{\mathrm{1}}}} \, = \, 0  \vdash  \ottkw{terminates}  :  \ottkw{Terminates}\ \, ( \,  \mathit{plus}  \, \mathit{x_{{\mathrm{2}}}} \, 0 \, ) \   \downarrow  
}{
 \Gamma  \vdash  \lambda \, \mathit{q} \, . \, \ottkw{terminates}  :   \Pi ^{  \downarrow  }  \mathit{q} \!:\! \mathit{x_{{\mathrm{1}}}} \, = \, 0 .  \ottkw{Terminates}\ \, ( \,  \mathit{plus}  \, \mathit{x_{{\mathrm{2}}}} \, 0 \, )  \   \downarrow  
}}
\]
For the successor case, we need to make a recursive call to the
theorem to show that the recursive call to the function terminates.
Below, let $\Gamma'$ be the extended environment
 $\Gamma \, , \, \mathit{z} \, : \, \ottkw{nat} \, , \, \mathit{q} \, : \, \mathit{x_{{\mathrm{1}}}} \, = \, \ottkw{Suc} \, \mathit{z}$ and 
$(*)$ be the derivation of $ \Gamma'  \vdash  \ottkw{join}  :   \mathit{plus}  \, \mathit{x_{{\mathrm{2}}}} \, ( \, \ottkw{Suc} \, \mathit{z} \, ) \, = \, \ottkw{Suc} \, ( \,  \mathit{plus}  \, \mathit{x_{{\mathrm{2}}}} \, \mathit{z} \, ) \   \downarrow  $.
Then, the derivation looks like:
\[
\prooftree
\prooftree
   {\prooftree
      {\prooftree      
        {\prooftree 
           \vdots
         \justifies
          \Gamma'  \vdash   \mathit{plus}  \, \mathit{x_{{\mathrm{2}}}} \, \mathit{z}  :  \ottkw{nat} \   ?  
        \endprooftree} \quad
      {\prooftree
       \vdots
      \justifies
         \Gamma'  \vdash  \mathit{f} \, \mathit{z}  :  \ottkw{Terminates}\ \, ( \,  \mathit{plus}  \, \mathit{x_{{\mathrm{2}}}} \, \mathit{z} \, ) \   \downarrow  
      \endprooftree}
     \justifies
      \Gamma'  \vdash   \mathit{plus}  \, \mathit{x_{{\mathrm{2}}}} \, \mathit{z}  :  \ottkw{nat} \   \downarrow   
    \endprooftree}
  \justifies
   \Gamma'  \vdash  \ottkw{Suc} \, ( \,  \mathit{plus}  \, \mathit{x_{{\mathrm{2}}}} \, \mathit{z} \, )  :  \ottkw{nat} \   \downarrow  
  \endprooftree}
  \justifies
   \Gamma'  \vdash  \ottkw{terminates}  :  \ottkw{Terminates}\ \, ( \, \ottkw{Suc} \, ( \,  \mathit{plus}  \, \mathit{x_{{\mathrm{2}}}} \, \mathit{z} \, ) \, ) \   \downarrow  
\endprooftree
  (*)
\justifies
 \Gamma'  \vdash  \ottkw{terminates}  :  \ottkw{Terminates}\ \, ( \,  \mathit{plus}  \, \mathit{x_{{\mathrm{2}}}} \, ( \, \ottkw{Suc} \, \mathit{z} \, ) \, ) \   \downarrow  
\endprooftree
\]

\paragraph{First-class termination proofs} Recursive functions can
also call helper functions in their definitions, passing off the
recursive term and a proof that the recursive call will terminate. 
For example, suppose there is some function $h$ that takes 
an argument, a (general) function to call on that argument, and a
proof that the call terminates.
\[ h :  \Pi ^{  \downarrow  }  \mathit{x} \!:\! \ottkw{nat} .   \Pi ^{  \downarrow  }  \mathit{f} \!:\!  \Pi ^{  ?  }  \mathit{x} \!:\! \ottkw{nat} .  \ottkw{nat}  .   \Pi ^{  \downarrow  }  \mathit{p} \!:\! \ottkw{Terminates}\ \, ( \, \mathit{f} \, \mathit{x} \, ) .  \ottkw{nat}    \]
For example, h may just apply $\mathit{f}$ to $\mathit{x}$ and use a
termination cast to show the effect total. We can use $\mathit{h}$ 
in the definition of a total recursive
function, even if we do not know its definition. (Let $\Gamma$ be 
a context which contains the above binding for $\mathit{h}$.)
\[  \Gamma  \vdash   \ottkw{rec} \  \mathit{f} ( \mathit{x} ) =  ( \,  \ottkw{case} \  \mathit{x} \  ( \, \lambda \, \mathit{q} \, . \, 0 \, ) \  ( \, \lambda \, \mathit{z} \, . \, \lambda \, \mathit{q} \, . \, \mathit{h} \, \mathit{z} \, \mathit{f} \, \ottkw{terminates} \, )  \, ) \, \ottkw{join}   :   \Pi ^{  \downarrow  }  \mathit{x} \!:\! \ottkw{nat} .  \ottkw{nat}  \   \downarrow   \]

Note that in this example, we use $\ottkw{terminates}$ as the proof that
$\mathit{f} \, \mathit{z}$ terminates. Although \ottdrulename{T\_RecNat} introduces the 
variable $\mathit{p}$, of type $ \Pi ^{  \downarrow  }  \mathit{z} \!:\! \ottkw{nat} .   \Pi ^{  \downarrow  }  \mathit{q} \!:\! \mathit{z} \, = \, \ottkw{Suc} \, \mathit{z} .  \ottkw{Terminates}\ \, ( \, \mathit{f} \, \mathit{z} \, )  $, we cannot pass $\mathit{p} \, \mathit{z} \, \mathit{q}$ as the termination proof to $\mathit{h}$ because
$\mathit{p}$ cannot be mentioned in the term. However, the proof term
$\ottkw{terminates}$ works instead, as shown by the following derivation.
(Let $\Gamma'$ be the context in the successor case, i.e. $\Gamma$
extended with bindings for $\mathit{x}$, $\mathit{f}$, $\mathit{p}$, $\mathit{z}$ and $\mathit{q}$.)
\[
\prooftree
  \prooftree
      \prooftree
         \vdots
      \justifies
          \Gamma'  \vdash  \mathit{p} \, \mathit{z} \, \mathit{q}  :  \ottkw{Terminates}\ \, ( \, \mathit{f} \, \mathit{z} \, ) \   \downarrow  
      \endprooftree
    \quad
      \prooftree
          \vdots
      \justifies
         \Gamma'  \vdash  \mathit{f} \, \mathit{z}  :  \ottkw{nat} \   ?  
      \endprooftree
  \justifies
    \Gamma'  \vdash  \mathit{f} \, \mathit{z}  :  \ottkw{nat} \   \downarrow  
  \using
   \ottdrulename{T\_Reflect}
  \endprooftree
\justifies
   \Gamma'  \vdash  \ottkw{terminates}  :  \ottkw{Terminates}\ \, ( \, \mathit{f} \, \mathit{z} \, ) \   \downarrow  
 \using
  \ottdrulename{T\_Reify}
\endprooftree
\]

\paragraph{Natural number division}

Finally, we demonstrate a function that requires a course-of-values
argument to show termination: natural number division. The general
problem is that division calls itself recursively on a number that is
smaller, but is not the direct predecessor of the argument. To show
that this function terminates, we do structural recursion on an upper
bound of the dividend instead of the dividend itself.  (Note that we
could also define division as a possibly partial function, without this
extra upper-bound argument, and separately write a proof that 
states that division is a total function.) 
The type we use for division is:
\[
 \mathit{div}  :  \Pi ^{  \downarrow  }  \mathit{z} \!:\! \ottkw{nat} .   \Pi ^{  \downarrow  }  \mathit{x} \!:\! \ottkw{nat} .   \Pi ^{  \downarrow  }  \mathit{x'} \!:\! \ottkw{nat} .   \Pi ^{  \downarrow  }  \mathit{u} \!:\! ( \,  \mathit{lte}  \, \mathit{x'} \, \mathit{x} \, ) \, = \, \ottkw{true} .  \ottkw{nat}    
\]
where $\mathit{z}$ is the divisor, $\mathit{x'}$ is the dividend, $\mathit{x}$ is 
an upper bound of the dividend, and $ \mathit{lte} $ is a function that 
determines if the first number is ``less-than-or-equal'' the second.
We have been parsimonious in
omitting a boolean type, so we use $0$ and $\ottkw{Suc} \, 0$ for
$\ottkw{false}$ and $\ottkw{true}$, respectively in the result of $ \mathit{lte} $. Therefore, we define
\[ lte \defeq   \ottkw{rec} \  \mathit{f} ( \mathit{x} ) =  \lambda \, \mathit{u} \, . \,  \ottkw{case} \  \mathit{x} \  ( \, \ottkw{Suc} \, 0 \, ) \  ( \, \lambda \, \mathit{x'} \, . \,  \ottkw{case} \  \mathit{u} \  0 \  ( \, \mathit{f} \, \mathit{x'} \, )  \, )   \]
and show
\[   \cdot   \vdash   \mathit{lte}   :   \Pi ^{  ?  }  \mathit{x} \!:\! \ottkw{nat} .   \Pi ^{  ?  }  \mathit{x'} \!:\! \ottkw{nat} .  \ottkw{nat}   \   \downarrow   \]
\noindent Note that we are considering $ \mathit{lte} $ as a possibly partial
function; nothing is harmed by not requiring it to be total.  We also
define cut-off subtraction as a total function $ \mathit{minus} $ of type
$ \Pi ^{  \downarrow  }  \mathit{x} \!:\! \ottkw{nat} .   \Pi ^{  \downarrow  }  \mathit{x'} \!:\! \ottkw{nat} .  \ottkw{nat}  $ (details omitted).
The code for division is then:
\[
\textit{div}\ \ \defeq\ \ 
\begin{array}[t]{ll}
 \lambda  \mathit{z} . (( \ottkw{case} \ \mathit{z}  \\
\quad\quad                 ( \, \lambda \, \mathit{q} \, . \, \lambda \, \mathit{x} \, . \, \lambda \, \mathit{x'} \, . \, \lambda \, \mathit{u} \, . \, 0 \, )  \\
\quad\quad                 ( \, \lambda \, \mathit{z'} \, . \, \lambda \, \mathit{q} \, . \,  \ottkw{rec} \  \mathit{f} ( \mathit{x} ) =  \lambda \, \mathit{x'} \, . \, \lambda \, \mathit{u} \, . \, ( \, ( \,  \ottkw{case} \  ( \,  \mathit{lte}  \, ( \, \ottkw{Suc} \, \mathit{x} \, ) \, \mathit{z} \, ) \  \ottnt{t_{{\mathrm{1}}}} \  ( \, \lambda \, \mathit{z''} \, . \, \lambda \, \mathit{q'} \, . \, 0 \, )  \, ) \, \ottkw{join} \, )  \, )) \\
\quad\ \ \ottkw{join})
\end{array}
\] 
\noindent We handle the case of division by $0$ up front, obtaining an
assumption $\mathit{q} : \mathit{z} \, = \, \ottkw{Suc} \, \mathit{z'}$ when the divisor is not zero. 
 Next, we case split on whether or not the bound
$x$ is strictly less than $z$; that is, $ \mathit{lte}  \, ( \, \ottkw{Suc} \, \mathit{x} \, ) \, \mathit{z}$.  If so, we use the term $\lambda \, \mathit{z''} \, . \, \lambda \, \mathit{q'} \, . \, 0$ of type
\[
 \Pi ^{  \downarrow  }  \mathit{z''} \!:\! \ottkw{nat} .   \Pi ^{  \downarrow  }  \mathit{q'} \!:\!  \mathit{lte}  \, ( \, \ottkw{Suc} \, \mathit{x} \, ) \, \mathit{z} \, = \, ( \, \ottkw{Suc} \, \mathit{z''} \, ) .  \ottkw{nat}  
\]
\noindent Then the quotient
is $0$.  If not, we use the
term $\ottnt{t_{{\mathrm{1}}}}$, of type $ \Pi ^{  \downarrow  }  \mathit{q'} \!:\! ( \,  \mathit{lte}  \, ( \, \ottkw{Suc} \, \mathit{x} \, ) \, \mathit{z} \, = \, 0 \, ) .  \ottkw{nat} $, which is (with $\ottnt{t_{{\mathrm{2}}}}$ discussed below):
\[
\ottnt{t_{{\mathrm{1}}}} \ \ \defeq \ \ \lambda \, \mathit{q'} \, . \, ( \, \ottkw{Suc} \, ( \, \mathit{f} \, ( \,  \mathit{pred}  \, \mathit{x} \, ) \, ( \,  \mathit{minus}  \, \mathit{x'} \, \mathit{z} \, ) \, \ottnt{t_{{\mathrm{2}}}} \, ) \, )
\]
\noindent In this case, we are decreasing our bound on the dividend by
one, and then using a termination cast to show that $\mathit{f} \, ( \,  \mathit{pred}  \, \mathit{x} \, )$
is terminating.  Here, we define $ \mathit{pred} $ as just $\lambda \, \mathit{x} \, . \,  \ottkw{case} \  \mathit{x} \  0 \  \lambda \, \mathit{x'} \, . \, \mathit{x'} $.  Of course, since this is the implicit language,
the termination cast does not appear in the term itself.  To apply the
termination cast, we must use the implicit assumption $p$ telling us
that $f$ terminates on the predecessor of $\mathit{x}$.  We can prove that
$ \ottkw{case} \  \mathit{x} \  0 \  \lambda \, \mathit{x'} \, . \, \mathit{x'} $ is the predecessor of $\mathit{x}$ in this case,
because the assumptions $\mathit{q} : \mathit{z} \, = \, ( \, \ottkw{Suc} \, \mathit{z'} \, )$ and 
$\mathit{q'} :  \mathit{lte}  \, ( \, \ottkw{Suc} \, \mathit{x} \, ) \, \mathit{z} \, = \, \ottkw{false}$
 show that $\mathit{x}$ is non-zero:
\noindent Intuitively, $\mathit{q'}$ implies that $x$ is greater than or
equal to $z$, which we know is non-zero by $\mathit{q}$.  The term $\ottnt{t_{{\mathrm{2}}}}$ is a proof that $ \mathit{minus}  \, \mathit{x'} \, \mathit{z}$ is less than or equal to the
predecessor of the bound, $ \ottkw{case} \  \mathit{x} \  0 \  \lambda \, \mathit{x'} \, . \, \mathit{x'} $.  In fact, $\ottkw{join}$ will serve for
$\ottnt{t_{{\mathrm{2}}}}$ because the desired equation is provable from the assumptions.

\section{A Logical Semantics for $\Teqt$}
\label{sec:consistency}

In this section, we give a semantics for $\Teqt$ in terms of a simple
constructive logic called $W'$. This semantics informs our design of
\Teqt and can potentially be used as part of a consistency proof for
\Teqt.  The theory $W'$ is reminiscent of Feferman's theory $W$ (see,
for example, Chapter 13 of~\cite{feferman98}).  $W$ is a classical
second-order theory of general-recursive functions, classified by
class terms which correspond to simple types.  $W$ supports
quantification over class terms, and quantification over defined
individual terms.  It is defined in Beeson's Logic of Partial Terms, a
logic designed for reasoning about definedness in the presence of
partial functions~\cite{B85}.  $W$ includes a relatively weak form of
natural-number induction.  Indeed, $W$ is conservative over Peano
Arithmetic.

\subsection{The theory $W'$}
\label{sec:wp}

\vspace{-2pt}
Figure~\ref{fig:syntaxwp} gives the syntax for sorts $\ottnt{A}$ (which
are just simple types) and formulas $\ottnt{F}$ for the theory $W'$; as
well as typing contexts $\Sigma$ and contexts $\ottnt{H}$ for logical
assumptions.  Terms $t$ are just as for (implicit) $\Teqt$, except
without $\ottkw{contra}$, $\ottkw{terminates}$, and $\ottkw{join}$.
Figure~\ref{fig:wp} gives the proof rules for the theory $W'$.  The
form of judgments is $\Sigma \, ; \, \ottnt{H} \, \vdash \, \ottnt{F}$.  This expresses that formula
$\ottnt{F}$ holds under the assumed formulas in $\ottnt{H}$.  $\Sigma$ is a
typing context declaring free term-level variables occurring in
$\ottnt{H}$ and $\ottnt{F}$.

$W'$ is similar in spirit to Feferman's $W$, but differs in a number
of details.  First, $W$ is a two-sorted theory: there is a sort for
individual terms, and one for class terms.  To express that term $t$
is in class $C$, theory $W$ uses an atomic formula $t\in C$.  Our
theory $W'$, in contrast, is a multi-sorted first-order logic, with
one sort for every simple type.  So $W'$ does not make use of a
predicate symbol to express that a term has a sort.  We only insist
that terms are well-sorted when instantiating quantifiers.  This is
apparent in the rule $\ottdrulename{Pv\_Alle}$, which depends on a
simple typing judgment for $W'$.  The rules for this typing judgment
may be found in 
\ifTR
the appendix (Section~\ref{sec:wptp}).
\else
the companion technical report \cite{teqtTR}.
\fi
Well-formedness of equations does not require well-sortedness of the
terms in $W'$ (as also in $W$).  Also, we have no reason at the moment
to include non-constructive reasoning in $W'$, so we define it using
principles of intuitionistic logic only.

A few more words on the proof principles of $W'$ are warranted. The
$\ottdrulename{Pv\_OpSem}$ equates terms $t$ and $t'$ iff $t\leadsto^*
t'$.  Thanks to the $\ottdrulename{Pv\_Subst}$ rule, symmetry and
transitivity of equality can be derived in a standard way.  We do not
require quantifiers to be instantiated by only terminating terms.
This means that for induction principles, we must state explicitly
that the terms in question are terminating.  We include a principle
$\ottdrulename{Pv\_CompInd}$ of computational induction, on the
structure of a terminating computation.  That is, if we know that an
application of a recursive function is terminating, we can prove a
property of such an application by assuming it is true for recursive
calls, and showing it is true for an outer arbitrary call of the
function.  Note that the assumption of termination of the application
of the recursive function is essential: without it, we could prove
diverging terms terminate.  We also include a principle
$\ottdrulename{Pv\_TermInv}$ of computational inversion, which allows
us to conclude $\ottkw{Terminates}\ \, \ottnt{t}$ from $\ottkw{Terminates}\ \, {\cal C} \, [ \, \ottnt{t} \, ]$.
Interestingly, even without the inversion rule of $\Teqt$, the
theorem we prove below would make heavy use of computational
inversion.  In a classical theory like $W$, this principle may well be
derivable from the other axioms.  Here, it does not seem to be.

\begin{figure}
\[
\begin{array}{rcl}
\ottnt{A} & ::= & \ottkw{nat} \alt  \ottnt{A}  \to  \ottnt{A'} \\
\ottnt{F} & ::= & \ottkw{True} \alt  \forall  \mathit{x}  :  \ottnt{A}  .  \ottnt{F}  \alt  \ottnt{F}  \Rightarrow  \ottnt{F'}  \alt  \ottnt{F}  \wedge  \ottnt{F'}  \alt \ottkw{Terminates}\ \, \ottnt{t} \alt \ottnt{t} \, = \, \ottnt{t'} \\
\Sigma & ::= &  \cdot  \alt \Sigma \, , \, \mathit{x} \, : \, \ottnt{A} \\
\ottnt{H} & ::= &  \cdot  \alt \ottnt{H} \, , \, \ottnt{F} \\
\end{array}
\]
\caption{Simple types, formulas, typing contexts, and assumption contexts of $W'$}
\label{fig:syntaxwp}
\end{figure}

\begin{figure}
\[
\begin{array}{c}
\begin{array}{ll}
\ottdrulePvXXAssume{} &
\ottdrulePvXXAlli{} \\ \\
\ottdrulePvXXAlle{} &
\ottdrulePvXXImpi{}\\ \\
\ottdrulePvXXImpe{} &
\ottdrulePvXXAndi{}\\ \\
\ottdrulePvXXAndeOne{} &
\ottdrulePvXXAndeTwo{}\\ \\
\ottdrulePvXXTruei{} &
\ottdrulePvXXContra{}\\ \\
\ottdrulePvXXOpSem{} &
\ottdrulePvXXSubst{} \\ \\
\ottdrulePvXXTermZero{} &
\ottdrulePvXXTermS{}\\ \\
\ottdrulePvXXTermAbs{} &
\ottdrulePvXXTermRec{}\\ \\
\ottdrulePvXXTermInv{} &
\ottdrulePvXXNotTermAbort{}
\end{array} \\ \\
\ottdrulePvXXInd{} \\ \\
\ottdrulePvXXCompInd{}\\ \\
\end{array}
\]
\caption{Theory $W'$}
\label{fig:wp}
\end{figure}

\vspace{-2pt}
\paragraph{Computational translation of terms}

\begin{figure}
\[
\begin{array}{lllllll}
\interp{\mathit{x}}^C & = & \mathit{x} & \ \ \ &
\interp{\ottnt{t} \, \ottnt{t'}}^C & = & \interp{\ottnt{t}}^C\ \interp{\ottnt{t'}}^C \\
\interp{\lambda \, \mathit{x} \, . \, \ottnt{t}}^C & = & \lambda x.\,\interp{\ottnt{t}}^C & \ \ \ &
\interp{0}^C & = & 0 \\
\interp{\ottkw{Suc} \, \ottnt{t}}^C & = & \ottkw{S}\ \interp{\ottnt{t}}^C & \ \ \ &
\interp{\ottkw{join}}^C & = & 0 \\
\interp{\ottkw{terminates}}^C & = & 0 & \ \ \ &
\interp{\ottkw{contra}}^C & = & 0 \\
\interp{\ottkw{abort}}^C & = & \ottkw{abort} & \ \ \ &\
\interp{ \ottkw{rec} \  \mathit{f} ( \mathit{x} ) =  \ottnt{t} }^C & = & \ottkw{rec}\ \mathit{f}(\mathit{x}). \interp{\ottnt{t}}^C \\
\interp{ \ottkw{case} \  \ottnt{t} \  \ottnt{t'} \  \ottnt{t''} }^C & = & \ottkw{C}\ \interp{\ottnt{t}}^C\ \interp{\ottnt{t'}}^C\ \interp{\ottnt{t''}}^C & \ \ \ &\ &\ &\ 
\end{array}
\]
\caption{Computational translation of terms}
\label{fig:compterm}
\end{figure}

Figure~\ref{fig:compterm} defines what we will refer to as the
computational translation of $\Teqt$ terms (the ``C'' is for
computational).  This translation, which is almost trivial, just maps
logical terms $\ottkw{join}$, $\ottkw{terminates}$, and $\ottkw{contra}$
to $0$.

\vspace{-2pt}
\paragraph{Translation of types}
Next, given $\Teqt$ type $T$, we define $\interp{T}^C$ and
$\interp{T}^L$.  The ``L'' is for logical translation.  This
$\interp{T}^C$ is a sort $\ottnt{A}$, and $\interp{T}^L$ is a predicate on
translated terms.  Recall that the syntax for such types and for the
formulas $\ottnt{F}$ used in such predicates is defined in
Figure~\ref{fig:syntaxwp} above. The definition of the interpretations
is then given in Figure~\ref{fig:interptp}.  Note that one can confirm
the well-foundedness of this definition by expanding the definition of
$\interp{\ottnt{T}}^L_\theta$, a convenient abbreviation, wherever it is
used.

\begin{figure}
\begin{minipage}[b]{0.4\linewidth}
\begin{eqnarray*}
\interp{\ottkw{nat}}^C & = & \ottkw{nat} \\
\interp{ \Pi ^{ \theta }  \mathit{x} \!:\! \ottnt{T} .  \ottnt{T'} }^C & = & \interp{T} \to \interp{T'} \\
\interp{\ottnt{t} \, = \, \ottnt{t'}}^C & = & \ottkw{nat} \\
\interp{\ottkw{Terminates}\ \, \ottnt{t}}^C & = & \ottkw{nat}
\end{eqnarray*}
\end{minipage}
\begin{minipage}[b]{0.6\linewidth}
\begin{eqnarray*}
\interp{\ottkw{nat}}^L\ t & = & \ottkw{True} \\
\interp{ \Pi ^{ \theta }  \mathit{x} \!:\! \ottnt{T} .  \ottnt{T'} }^L\ t & = & \forall \mathit{x} : \interp{T}^C. \interp{T}^L_\downarrow\ x\ \Rightarrow \interp{T'}^L_\theta\ (t\ x) \\
\interp{\ottnt{t_{{\mathrm{1}}}} \, = \, \ottnt{t_{{\mathrm{2}}}}}^L\ t & = & \interp{\ottnt{t_{{\mathrm{1}}}}}^C = \interp{\ottnt{t_{{\mathrm{2}}}}}^C \\
\interp{\ottkw{Terminates}\ \, \ottnt{t'}}^L\ t & = & \ottkw{Terminates}\ \interp{\ottnt{t'}}^C 
\end{eqnarray*}
\end{minipage}
\begin{eqnarray*}
\interp{\ottnt{T}}^L_\downarrow\ t & = & \ottkw{Terminates}\ \ottnt{t}\ \wedge\ \interp{\ottnt{T}}^L\ \ottnt{t} \\
\interp{\ottnt{T}}^L_?\ t & = & \ottkw{Terminates}\ \ottnt{t}\ \Rightarrow\ \interp{\ottnt{T}}^L\ \ottnt{t} 
\end{eqnarray*}
\caption{Interpretation of types}
\label{fig:interptp}
\end{figure}

\subsection{Examples}

\noindent \textbf{Example 1.} If we consider the type $ \Pi ^{  \downarrow  }  \mathit{x_{{\mathrm{1}}}} \!:\! \ottkw{nat} .   \Pi ^{  \downarrow  }  \mathit{x_{{\mathrm{2}}}} \!:\! \ottkw{nat} .  \ottkw{nat}  $, we will get the following.
Note that the assumptions below that variables terminate reflect the
call-by-value nature of the language.  A translation for a
call-by-name language would presumably not include such assumptions.
\begin{eqnarray*}
\interp{ \Pi ^{  \downarrow  }  \mathit{x_{{\mathrm{1}}}} \!:\! \ottkw{nat} .   \Pi ^{  \downarrow  }  \mathit{x_{{\mathrm{2}}}} \!:\! \ottkw{nat} .  \ottkw{nat}  }^C & = &  \ottkw{nat}  \to  ( \,  \ottkw{nat}  \to  \ottkw{nat}  \, ) \\ 
\interp{ \Pi ^{  \downarrow  }  \mathit{x_{{\mathrm{1}}}} \!:\! \ottkw{nat} .   \Pi ^{  \downarrow  }  \mathit{x_{{\mathrm{2}}}} \!:\! \ottkw{nat} .  \ottkw{nat}  }^L\ \textit{plus} & = & 
\forall x_1 : \ottkw{nat} .\, \ottkw{Terminates}\ x_1 \wedge \ottkw{True} \Rightarrow \ottkw{Terminates}\ (\textit{plus}\ x_1)\ \wedge\ \\
\ &\ & \forall x_2 : \ottkw{nat} .\, \ottkw{Terminates}\ x_2 \wedge \ottkw{True} \Rightarrow \ottkw{Terminates}\ (\textit{plus}\ x_1\ x_2)\\
\ &\ & \ \ \ \ \ \wedge\ \ottkw{True}
\end{eqnarray*}

\noindent \textbf{Example 2 (higher-order, total).} If we wanted to type a
function \textit{iter} which iterates a terminating function $\mathit{x_{{\mathrm{1}}}}$,
starting from $\mathit{x_{{\mathrm{2}}}}$, and does this iteration $\mathit{x_{{\mathrm{3}}}}$ times, we
might use the type: \[  \Pi ^{  \downarrow  }  \mathit{x_{{\mathrm{1}}}} \!:\!  \Pi ^{  \downarrow  }  \mathit{x} \!:\! \ottkw{nat} .  \ottkw{nat}  .   \Pi ^{  \downarrow  }  \mathit{x_{{\mathrm{2}}}} \!:\! \ottkw{nat} .   \Pi ^{  \downarrow  }  \mathit{x_{{\mathrm{3}}}} \!:\! \ottkw{nat} .  \ottkw{nat}   . \]  For this type (call it
$T$ for brevity), we will get the following translations:
\begin{eqnarray*}
\interp{T}^C & = &  ( \,  \ottkw{nat}  \to  \ottkw{nat}  \, )  \to  ( \,  \ottkw{nat}  \to  ( \,  \ottkw{nat}  \to  \ottkw{nat}  \, )  \, ) \\ 
\interp{T}^L\ \textit{iter} & = & 
\forall x_1 : \ottkw{nat} \to \ottkw{nat}.\, \ottkw{Terminates}\ x_1 \wedge\\
\ &\ &\ \ \ \ \ \ \ \ \ \ 
 (\forall x : \ottkw{nat}. \ottkw{Terminates}\ \, \mathit{x}\ \wedge\ \ottkw{True}\Rightarrow\ottkw{Terminates}\ (x_1\ x)\ \wedge\ \ottkw{True})\ \Rightarrow\\ 
\ &\ & \ \ \ \ \ \ \ \ \ \ \ \ \ \ \ \ \ \ \ \ \ \ \ \ \ \ \ \ \ \ \ \ \ \ \ \ \ \ \ \ \ \ \ \ \ \ \ \ \ \ \ \ \ \ \ \ \
\ottkw{Terminates}\ (\textit{\textit{iter}}\ x_1)\ \wedge\ \\
\ &\ & \forall x_2 : \ottkw{nat} .\, \ottkw{Terminates}\ x_2 \wedge \ottkw{True} \Rightarrow \ottkw{Terminates}\ (\textit{\textit{iter}}\ x_1\ x_2)\ \wedge\\
\ &\ & \forall x_3 : \ottkw{nat} .\, \ottkw{Terminates}\ x_3 \wedge \ottkw{True} \Rightarrow \ottkw{Terminates}\ (\textit{\textit{iter}}\ x_1\ x_2\ x_3)\ \wedge\ \ottkw{True}
\end{eqnarray*}

\noindent Notice that in this case, the logical interpretation
$\interp{T}^L$ includes a hypothesis that the function $\mathit{x_{{\mathrm{1}}}}$ is
terminating.  This corresponds to the fact that $\mathit{x_{{\mathrm{1}}}}$ has type
$ \Pi ^{  \downarrow  }  \mathit{x} \!:\! \ottkw{nat} .  \ottkw{nat} $ in the original $\Teqt$ type.

\ 

\noindent \textbf{Example 3 (higher-order, partial).} If we wanted to
type a different version of \textit{iter} which, when given a
general-recursive function $\mathit{x_{{\mathrm{1}}}}$ and a starting value $\mathit{x_{{\mathrm{2}}}}$,
returns a general-recursive function taking input $\mathit{x_{{\mathrm{3}}}}$ and
iterating $\mathit{x_{{\mathrm{1}}}}$ $\mathit{x_{{\mathrm{3}}}}$ times starting from $\mathit{x_{{\mathrm{2}}}}$, we might use
the type: \[  \Pi ^{  \downarrow  }  \mathit{x_{{\mathrm{1}}}} \!:\!  \Pi ^{  ?  }  \mathit{x} \!:\! \ottkw{nat} .  \ottkw{nat}  .   \Pi ^{  \downarrow  }  \mathit{x_{{\mathrm{2}}}} \!:\! \ottkw{nat} .   \Pi ^{  ?  }  \mathit{x_{{\mathrm{3}}}} \!:\! \ottkw{nat} .  \ottkw{nat}   . \]  For this type (call it $T$),
we will get the following logical translation:
\begin{eqnarray*}
\interp{T}^L\ \textit{iter} & = & 
\forall x_1 : \ottkw{nat} \to \ottkw{nat}.\, \ottkw{Terminates}\ x_1 \wedge\\
\ &\ & \ \ \ \ \ \ \ \ \ \ (\forall x : \ottkw{nat}. \ottkw{Terminates}\ \, \mathit{x}\ \wedge\ \ottkw{True}\Rightarrow\ottkw{Terminates}\ (x_1\ x)\ \Rightarrow\ \ottkw{True})\ \Rightarrow\\ 
\ &\ & \ \ \ \ \ \ \ \ \ \ \ \ \ \ \ \ \ \ \ \ \ \ \ \ \ \ \ \ \ \ \ \ \ \ \ \ \ \ \ \ \ \ \ \ \ \ \ \ \ \ \ \ \ \ \ \ \ 
\ottkw{Terminates}\ (\textit{\textit{iter}}\ x_1)\ \wedge\ \\
\ &\ & \forall x_2 : \ottkw{nat} .\, \ottkw{Terminates}\ x_2 \wedge \ottkw{True} \Rightarrow \ottkw{Terminates}\ (\textit{\textit{iter}}\ x_1\ x_2)\ \wedge\\
\ &\ & \forall x_3 : \ottkw{nat} .\, \ottkw{Terminates}\ x_3 \wedge \ottkw{True} \Rightarrow \ottkw{Terminates}\ (\textit{\textit{iter}}\ x_1\ x_2\ x_3)\ \Rightarrow\ \ottkw{True}
\end{eqnarray*}

\subsection{Translation of contexts}

Figure~\ref{fig:interpctxt} gives a similar 2-part translation of
typing contexts.  The translation $\interp{\cdot}^C$ produces a
simple-typing context $\Sigma$, while the translation
$\interp{\cdot}^L$ produces a logical context $\ottnt{H}$, which asserts,
for each variable $\mathit{x}$, that $\mathit{x}$ terminates and has the
property given by the $\interp{\cdot}^L$ translation of its type.

\begin{figure}
\[
\begin{array}[t]{lcl}
\interp{ \cdot }^C & = &  \cdot  \\
\interp{\Gamma \, , \, \mathit{x} \, : \, \ottnt{T}}^C & = & \interp{\Gamma}, \mathit{x} :
\interp{\ottnt{T}}^C\\\\
\end{array}
\qquad
\begin{array}[t]{lcl}
\interp{ \cdot }^L & = &  \cdot  \\
\interp{\Gamma \, , \, \mathit{x} \, : \, \ottnt{T}}^L & = & \interp{\Gamma},\, \interp{\ottnt{T}}^L_\downarrow\ \mathit{x}
\end{array}
\]
\vspace{-3ex}
\caption{Interpretation of contexts}
\label{fig:interpctxt}
\end{figure}

\subsection{Translation of typing judgments}

We are now in a position to state the main theorems of this paper.
The proofs are given in
\ifTR
the Appendix.  
\else
the companion technical report.
\fi
Theorem~\ref{thm:sndl} shows
that the logical translation of types is sound: the property expressed
by $\interp{\ottnt{T}}^L_\theta$ can indeed be proved to hold for the
translation $\interp{\ottnt{t}}^C$ of terms of type $T$.

\begin{thm}[Soundness of Computational Translation] 
\label{thm:sndc}
If $ \Gamma  \vdash  \ottnt{t}  :  \ottnt{T} \  \theta $, then $\interp{\Gamma}^C \vdash \interp{\ottnt{t}}^C : \interp{\ottnt{T}}^C$.
\end{thm}

\begin{thm}[Soundness of Logical Translation] 
\label{thm:sndl}
If\ \ $ \Gamma  \vdash  \ottnt{t}  :  \ottnt{T} \  \theta $, then $\interp{\Gamma}^C ; \interp{\Gamma}^L
\vdash \interp{\ottnt{T}}^L_\theta\ \interp{\ottnt{t}}^C$.
\end{thm}

\section{Related Work}
\label{sec:related}

\paragraph{Capretta's Partiality Monad}
Capretta~\cite{capretta05} gives an account of general recursion in
terms of a coinductive type constructor $(\cdot)^\nu$, and many
$\Teqt$ programs can be fairly mechanically translated into programs
using $(\cdot)^\nu$ by a translation similar to the the one described by
Wadler and Thiemann~\cite{wadler+03}. However, one interesting
difference is that $\Teqt$ functions can have a return type which
depends on a potentially nonterminating argument. It is not clear how
to represent this in a monadic framework.

For example, if we imagine a version of $\Teqt$ extended with option
types, and suppose we are given a decision procedure for equality of
$\ottkw{nat}$s and a partial function which computes the minimum zero of a
function:
\[
\begin{array}{l}
   \mathit{eqDec}  :  \Pi ^{  \downarrow  }  \mathit{x} \!:\! \ottkw{nat} .   \Pi ^{  \downarrow  }  \mathit{x'} \!:\! \ottkw{nat} .  \ottkw{Maybe} \, ( \, \mathit{x} \, = \, \mathit{x'} \, )   \\
   \mathit{minZero}  :  \Pi ^{  ?  }  \mathit{f} \!:\! ( \,  \Pi ^{  \downarrow  }  \mathit{x} \!:\! \ottkw{nat} .  \ottkw{nat}  \, ) .  \ottkw{nat} 
\end{array}
\]
Then we can easily compose these to make a function to test if two functions
have the same least zero:
\[
\begin{array}{l}
\lambda \, \mathit{f} \, . \, \lambda \, \mathit{f'} \, . \,  \mathit{eqDec}  \, ( \,  \mathit{minZero}  \, \mathit{f} \, ) \, ( \,  \mathit{minZero}  \, \mathit{f'} \, ) \\
 \qquad :  \Pi ^{  \downarrow  }  \mathit{f} \!:\! ( \,  \Pi ^{  \downarrow  }  \mathit{x} \!:\! \ottkw{nat} .  \ottkw{nat}  \, ) .   \Pi ^{  ?  }  \mathit{f'} \!:\! ( \,  \Pi ^{  \downarrow  }  \mathit{x} \!:\! \ottkw{nat} .  \ottkw{nat}  \, ) .  \ottkw{Maybe} \, ( \,  \mathit{minZero}  \, \mathit{f} \, = \,  \mathit{minZero}  \, \mathit{f'} \, )  
\end{array}
\]
However the naive translation of this into monadic form,
\[
 \lambda f . \lambda  f' . (\mathit{minZero}\ f)\ \ottkw{>\!\!>\!=}\ (\lambda m . (\mathit{minZero}\ f')\ \ottkw{>\!\!>\!=}\ (\lambda m' . \ottkw{return}\ (\mathit{eqDec}\ m\ m'))),
\]
is not well typed, since the monadic bind $\ottkw{>\!\!>\!=} : \forall A\
B. \, A^\nu  \rightarrow  (A \rightarrow B^\nu) \rightarrow B^\nu$ does not have a way to propagate the type dependency.

\paragraph{Other} Another approach, not depending on coinductive types, is explored by
Capretta and Bove, who define a special-purpose accessibility
predicate for each general-recursive function, and then define the
function by structural recursion on the proof of accessibility for the
function's input~\cite{bove+05}.  \textsc{ATS} and \textsc{Guru} both
separate the domains of proofs and programs, and can thus allow
general recursion without endangering logical
soundness~\cite{guru09,chenxi05}.  Systems like
Cayenne~\cite{Aug98}, $\Omega$\textsc{mega}~\cite{sheard06}.  and
\textsc{Concoqtion}~\cite{pasalic+07} support dependent types and
general recursion, but do not seek to identify a fragment of the term
language which is sound as a proof system (although
\textsc{Concoqtion} uses \textsc{Coq} proofs for reasoning about
type indices).

\section{Conclusion}

$\Teqt$ combines equality types and general recursion, using an effect
system to distinguish total from possibly partial terms.  Termination
casts are used to change the type system's view of the termination
behavior of a term.  Like other casts, termination casts have no
computational relevance and are erased in passing from the annotated
to the implicit type system.  We have given a logical semantics for
$\Teqt$ in terms of a multi-sorted first-order theory of
general-recursive functions.  Future work includes further
meta-theory, including type soundness for $\Teqt$ and further analysis
of the proposed theory $W'$; as well as incorporation of other typing
features, in particular polymorphism and large eliminations.  An
important further challenge is devising algorithms to reconstruct
annotations in simple cases or for common programming idioms.

\textbf{Acknowledgments.} Many thanks to the PAR 2010
reviewers for an exceptionally close reading and many constructive
criticisms. All
syntax definitions in this paper were typeset and type-checked with
the \textsc{Ott} tool~\cite{sewell+10}.
Thanks also to other members of the \textsc{Trellys} project,
especially Tim Sheard, for helpful conversations on these ideas.  This
work was partially supported by the U.S. National Science Foundation
under grants 0702545, 0910510 and 0910786.
  

\bibliographystyle{eptcs}
\bibliography{biblio}

\ifTR
\appendix

\section{Proof of Theorem~\ref{thm:sndc} (Soundness of Computational Translation)}

The proof is a routine induction on the assumed $\Teqt$ typing derivation, which
we include here for thoroughness:

\pfcase{T\_Var}{\ottdruleTXXVar{}} This case follows directly from the easily
proven fact that $ \Gamma ( \mathit{x} ) =  \ottnt{T} $ implies $\interp{\Gamma}^C(\mathit{x}) =
\interp{\ottnt{T}}^C$.  

\pfcase{T\_Join}{\ottdruleTXXJoin{}} The interpretation of the conclusion is
just an instance of the $\ottdrulename{STy\_Var}$ rule.  This is also true
for the rules $\ottdrulename{T\_Reify}$, $\ottdrulename{T\_Inv}$, and
$\ottdrulename{T\_Contra}$, so we omit cases for those rules.

\pfcase{T\_Conv}{\ottdruleTXXConv{}} By the IH we have $\interp{\Gamma}^C \vdash
\interp{\ottnt{t}}^C : \interp{[ \, \ottnt{t_{{\mathrm{2}}}} \, / \, \mathit{x} \, ] \, \ottnt{T}}^C$.  We omit the
straightforward proof that $\interp{[ \, \ottnt{t_{{\mathrm{2}}}} \, / \, \mathit{x} \, ] \, \ottnt{T}}^C =
\interp{\ottnt{T}}^C = \interp{[ \, \ottnt{t_{{\mathrm{1}}}} \, / \, \mathit{x} \, ] \, \ottnt{T}}^C$, for any $\ottnt{t_{{\mathrm{1}}}}$,
$\ottnt{t_{{\mathrm{2}}}}$, $\mathit{x}$, and $\ottnt{T}$.  So the fact we have from the first
premise is what is required for the conclusion.  The case for
$\ottdrulename{T\_Reflect}$ is similar, and so is omitted.

\pfcase{T\_Abs}{\ottdruleTXXAbs{}} By the IH we have $\interp{\Gamma \, , \, \mathit{x} \, : \, \ottnt{T'}}^C \vdash \interp{\ottnt{t}}^C : \interp{\ottnt{T}}^C$.  This is
equivalent to $\interp{\Gamma}^C , x : \interp{\ottnt{T'}}^C \vdash
\interp{\ottnt{t}}^C : \interp{\ottnt{T}}^C$, to which we can apply the simple
typing rule $\ottdrulename{STy\_Abs}$ to obtain $\interp{\Gamma}^C \vdash
\lambda \mathit{x} .\, \interp{\ottnt{t}}^C : \interp{\ottnt{T'}}^C\to
\interp{\ottnt{T}}^C$, which suffices by the definition of $\interp{\cdot}^C$.

\pfcase{T\_App}{\ottdruleTXXApp{}} By the IH and the definition of
$\interp{\cdot}^C$, we have $\interp{\Gamma}^C\vdash \interp{\ottnt{t}}^C :
\interp{\ottnt{T'}}^C\to\interp{\ottnt{T}}^C$ and also $\interp{\Gamma}^C\vdash
\interp{\ottnt{t'}}^C : \interp{\ottnt{T'}}^C$.  We may apply the simple
typing rule $\ottdrulename{STy\_App}$ to get $\interp{\Gamma}^C\vdash
\interp{\ottnt{t}}^C\ \interp{\ottnt{t'}}^C : \interp{\ottnt{T}}^C$, which
suffices, using again the definition of $\interp{\cdot}^C$, and also
the fact used above that $\interp{[ \, \ottnt{t'} \, / \, \mathit{x} \, ] \, \ottnt{T}}^C =
\interp{\ottnt{T}}^C$.

\pfcase{T\_Zero}{\ottdruleTXXZero{}} The desired conclusion is an instance of
$\ottdrulename{STy\_Zero}$.

\pfcase{T\_Suc}{\ottdruleTXXSuc{}} This case follows from the IH and then applying
$\ottdrulename{STy\_Suc}$.

\pfcase{T\_RecNat}{\ottdruleTXXRecNat{}} By the IH, we have:
\[
 \interp{\Gamma \, , \, \mathit{f} \, : \,  \Pi ^{  ?  }  \mathit{x} \!:\! \ottkw{nat} .  \ottnt{T}  \, , \, \mathit{x} \, : \, \ottkw{nat} \, , \, \mathit{p} \, : \,  \Pi ^{  \downarrow  }  \mathit{x_{{\mathrm{1}}}} \!:\! \ottkw{nat} .   \Pi ^{  \downarrow  }  \mathit{x_{{\mathrm{2}}}} \!:\! \mathit{x} \, = \, \ottkw{Suc} \, \mathit{x_{{\mathrm{1}}}} .  \ottkw{Terminates}\ \, ( \, \mathit{f} \, \mathit{x_{{\mathrm{1}}}} \, )  }^C \vdash \interp{\ottnt{t}}^C : \interp{\ottnt{T}}^C
\]
\noindent This is equivalent to:
\[
 \interp{\Gamma}^C, f : \ottkw{nat}\to\interp{\ottnt{T}}^C, x:\ottkw{nat}, p:\ottkw{nat}\to \ottkw{nat} \to \ottkw{nat} \vdash \interp{\ottnt{t}}^C : \interp{\ottnt{T}}^C
\]
\noindent We omit the straightforward proof that $\ottkw{fv}
\interp{\ottnt{t}}^C = \ottkw{fv} \ottnt{t}$, which gives us
$p\not\in\ottkw{fv} \interp{\ottnt{t}}^C$.  We also omit the
straightforward proof of Strengthening for our simply typed system,
which says $ \Sigma \, , \, \mathit{x} \, : \, \ottnt{A}  \vdash  \ottnt{t}  :  \ottnt{A'} $ implies $ \Sigma  \vdash  \ottnt{t}  :  \ottnt{A'} $ if $\mathit{x} \, \not\in \, \ottkw{fv} \, \ottnt{t}$.  Using this Strengthening property for the simply
typed system, we then have:
\[
 \interp{\Gamma}^C, f : \ottkw{nat}\to\interp{\ottnt{T}}^C, x:\ottkw{nat} \vdash \interp{\ottnt{t}}^C : \interp{\ottnt{T}}^C
\]
\noindent We may now just apply the rule $\ottdrulename{STy\_Rec}$, to
conclude the desired $\interp{\Gamma}^C \vdash \ottkw{rec}\, f\, ( x
)\, = \interp{\ottnt{t}}^C : \ottkw{nat}\to\interp{\ottnt{T}}^C$.  The case for
$\ottdrulename{T\_Rec}$ is the same as the last part of this case, and
so is omitted.

\pfcase{T\_Case}{\ottdruleTXXCase{}} By the IH and the definition of
$\interp{\cdot}^C$, we have:
\begin{itemize}
\item $\interp{\Gamma}^C \vdash \interp{t}^C : \ottkw{nat}$
\item $\interp{\Gamma}^C \vdash \interp{t'}^C : \interp{[0/x]\ottnt{T}}^C$
\item $\interp{\Gamma}^C \vdash \interp{t''}^C : \ottkw{nat}\to \interp{[Suc x'/x]\ottnt{T}}^C$
\end{itemize}
\noindent Using again the property mentioned above, that $\interp{[ \, \ottnt{t} \, / \, \mathit{x} \, ] \, \ottnt{T}}^C = \interp{\ottnt{T}}^C$, we may then apply the
rule $\ottdrulename{STy\_Case}$ to obtain the desired conclusion.

\pfcase{T\_Abort}{\ottdruleTXXAbort{}} The desired conclusion is just an
instance of $\ottdrulename{STy\_Abort}$.

\section{Proof of Theorem~\ref{thm:sndl} (Soundness of Logical Translation)}

We prove this by induction on the structure of the assumed derivation.
Note first that if the interpretation of a $\Teqt$ typing judgment
with effect $\downarrow$ holds -- that is, if we have
$\interp{\Gamma}^C ; \interp{\Gamma}^L \vdash
\ottkw{Terminates}\ \interp{t}^C\ \wedge\ \interp{\ottnt{T}}^L
\interp{\ottnt{t}}^C$ -- then we also have $\interp{\Gamma}^C ;
\interp{\Gamma}^L \vdash
\ottkw{Terminates}\ \interp{t}^C\ \Rightarrow\ \interp{\ottnt{T}}^L
\interp{\ottnt{t}}^C$, which is the interpretation of the similar $\Teqt$
typing judgment with effect $?$.  So in cases where we can prove the
interpretation of the judgment with $\downarrow$, we can omit the
proof of the interpretation of the judgment with $?$.

\pfcase{T\_Var}{\ottdruleTXXVar{}} This case follows directly from the fact
that the logical interpretation of the $\Teqt$ typing context $\Gamma$
must contain $\ottkw{Terminates}\ \mathit{x}$ and $\interp{\ottnt{T}}^L\ x$,
since $\Gamma(x) = T$.

\pfcase{T\_Join}{\ottdruleTXXJoin{}} From the fact that $\ottnt{t}
\leadsto^* \ottnt{t_{{\mathrm{0}}}}$ implies $\interp{\ottnt{t}}^C \leadsto^*
\interp{\ottnt{t_{{\mathrm{0}}}}}^C$ (we omit the easy proof), we have that
$\interp{\ottnt{t}}^C$ and $\interp{\ottnt{t'}}^C$ are joinable.  Our
equational theory allows us to prove that joinable terms are equal
(regardless of whether they are terminating or not).  Hence, we can
indeed prove the logical interpretation of the type in the conclusion,
namely $\interp{\ottnt{t}}^C = \interp{\ottnt{t'}}^C$.

\pfcase{T\_Conv}{\ottdruleTXXConv{}} By the IH we have:
\[
\interp{\Gamma}^C; \interp{\Gamma}^L \vdash \interp{[ \, \ottnt{t_{{\mathrm{2}}}} \, / \, \mathit{x} \, ] \, \ottnt{T}}^L_\theta\ \interp{\ottnt{t}}^C
\]
\noindent We omit the straightforward proof that 
\[
\interp{[ \, \ottnt{t_{{\mathrm{2}}}} \, / \, \mathit{x} \, ] \, \ottnt{T}}^L_\theta\ \interp{\ottnt{t}}^C= \ [ \interp{\ottnt{t_{{\mathrm{2}}}}}^C / x ] (\interp{\ottnt{T}}^L_\theta \ \interp{\ottnt{t}}^C)
\]
\noindent Using this fact, it suffices to prove the similar statement,
except with $\interp{\ottnt{t_{{\mathrm{1}}}}}^C$ in place of $\interp{\ottnt{t_{{\mathrm{2}}}}}^C$.  But
this follows by $\ottdrulename{Pv\_Subst}$, using the fact that
$\interp{\Gamma}^C; \interp{\Gamma}^L \vdash \interp{\ottnt{t_{{\mathrm{1}}}}}^C =
\interp{\ottnt{t_{{\mathrm{2}}}}}^C$.  We have this from the formula we get by the IH
for the second premise, noting that
\[
\interp{\ottnt{t_{{\mathrm{1}}}} \, = \, \ottnt{t_{{\mathrm{2}}}}}^L_\downarrow \ \interp{\ottnt{t'}}^C\ = \ \ottkw{Terminates}\ \interp{\ottnt{t'}}^C\ \wedge\ \interp{\ottnt{t_{{\mathrm{1}}}}}^C = \interp{\ottnt{t_{{\mathrm{2}}}}}^C
\]

\pfcase{T\_Reflect}{\ottdruleTXXReflect{}} If $\theta = \theta'$, the desired
result follows immediately from the IH applied to the first premise.
If $\theta \neq \theta'$ but $\theta = \downarrow$, we have already
observed above that we can obtain the logical translation of a $\Teqt$
typing judgment with effect $?$ if we have the similar translation
with effect $\downarrow$.  So it suffices to consider just the case
where $\theta = ?$ but $\theta' = \downarrow$.  By the IH for the
second premise, we have:
\[
\interp{\Gamma}^C ; \interp{\Gamma}^L \vdash \ottkw{Terminates}\ \interp{\ottnt{t'}}^C\ \wedge\ \ottkw{Terminates}\ \interp{\ottnt{t}}^C
\]
\noindent The second conjunct of this is exactly what we need to obtain the desired
conclusion from what we get from the IH applied to the first premise, which is:
\[
\interp{\Gamma}^C ; \interp{\Gamma}^L \vdash \ottkw{Terminates}\ \interp{\ottnt{t}}^C\ \Rightarrow\ \interp{\ottnt{T}}^L\ \interp{\ottnt{t}}^C
\]

\pfcase{T\_Reify}{\ottdruleTXXReify{}} The IH for the premise is:
\[
\interp{\Gamma}^C ; \interp{\Gamma}^L \vdash \ottkw{Terminates}\ \interp{\ottnt{t}}^C\ \wedge\ \interp{\ottnt{T}}^L\ \interp{\ottnt{t}}^C
\]
\noindent From this by $\ottdrulename{Pv\_Ande1}$ we obtain the
translation of the conclusion, using also the axiom
$\ottdrulename{Pv\_Terminates0}$ (to show
$\ottkw{Terminates}\ \interp{\ottkw{terminates}}^C$).

\pfcase{T\_CtxTerm}{\ottdruleTXXCtxTerm{}} It is sufficient to show
$\ottkw{Terminates}\ \interp{{\cal C}}^C[\interp{\ottnt{t'}}^C]\ \Rightarrow\ \ottkw{Terminates}\ \interp{\ottnt{t'}}^C$,
where $\interp{{\cal C}}^C$ is the context determined by the obvious
extension of $\interp{\cdot}^C$ from terms to contexts.  This formula
easily follows using $\ottdrulename{Pv\_TermInv}$.

\pfcase{T\_Abs}{\ottdruleTXXAbs{}} Applying the IH to the first premise gives us:
\[
\interp{\Gamma}^C, x:\interp{\ottnt{T'}}^C ; \interp{\Gamma}^L,\,\interp{\ottnt{T'}}^L_\downarrow\ x \vdash \interp{\ottnt{T}}^L_\theta \interp{\ottnt{t}}^C
\]
\noindent As remarked above, it suffices to prove the conclusion for when $\theta'=\downarrow$, since
this implies the case when $\theta'=?$.  So we must prove:
\[
\interp{\Gamma}^C ; \interp{\Gamma}^L \vdash \ottkw{Terminates}\ \, \lambda \, \mathit{x} \, . \, \ottnt{t} \wedge \forall \mathit{x} : \interp{\ottnt{T'}}^C.\ \interp{\ottnt{T'}}^L_\downarrow\ x\ \Rightarrow \interp{T}^L_\theta\ \interp{( \, \lambda \, \mathit{x} \, . \, \ottnt{t} \, ) \, \mathit{x}}^C
\]
\noindent The first conjunct is provable by
$\ottdrulename{Pv\_TermAbs}$.  The second follows easily from
the fact we obtained from the IH, by applying
$\ottdrulename{Pv\_Subst}$ with the equation $\ottnt{t} \, = \, ( \, \lambda \, \mathit{x} \, . \, \ottnt{t} \, ) \, \mathit{x}$,
which holds by $\ottdrulename{Pv\_OpSem}$ (and then using also
$\ottdrulename{Pv\_Alli}$ and $\ottdrulename{Pv\_Impi}$).

\pfcase{T\_App}{\ottdruleTXXApp{}} We first case-split on whether
$\theta=?$ or $\theta=\downarrow$.  If $\theta=?$, then by the IH, we have:
\begin{itemize}
\item $\interp{\Gamma}^C ; \interp{\Gamma}^L \vdash \ottkw{Terminates}\, \interp{\ottnt{t}}^C \Rightarrow \forall \mathit{x} : \interp{\ottnt{T'}}^C.\ \interp{\ottnt{T'}}^L_\downarrow\ x\ \Rightarrow \interp{\ottnt{T}}^L_\rho\ \interp{\ottnt{t}}^C\ x$
\item $\interp{\Gamma}^C ; \interp{\Gamma}^L \vdash \ottkw{Terminates}\, \interp{\ottnt{t'}}^C \Rightarrow \interp{\ottnt{T'}}^L\ \interp{\ottnt{t'}}^C$
\end{itemize}
\noindent We must prove:
\[
\interp{\Gamma}^C ; \interp{\Gamma}^L \vdash \ottkw{Terminates}\, \interp{\ottnt{t} \, \ottnt{t'}}^C \Rightarrow \interp{[ \, \ottnt{t'} \, / \, \mathit{x} \, ] \, \ottnt{T}}^L\ \interp{\ottnt{t} \, \ottnt{t'}}^C
\]
\noindent So (using $\ottdrulename{Pv\_Impi}$) assume
$\ottkw{Terminates}\, \interp{\ottnt{t} \, \ottnt{t'}}^C$, and prove $\interp{[ \, \ottnt{t'} \, / \, \mathit{x} \, ] \, \ottnt{T}}^L\ \interp{\ottnt{t} \, \ottnt{t'}}^C$.  By
$\ottdrulename{Pv\_TermInv}$, we have $\ottkw{Terminates}\,
\interp{\ottnt{t}}^C$ and $\ottkw{Terminates}\, \interp{\ottnt{t'}}^C$.
So from the facts we obtained above by the IH, we get (using $\ottdrulename{Pv\_Impe})$: 
\begin{itemize}
\item $\interp{\Gamma}^C ; \interp{\Gamma}^L \vdash \forall \mathit{x} : \interp{\ottnt{T'}}^C.\ \interp{\ottnt{T'}}^L_\downarrow\ x\ \Rightarrow \interp{\ottnt{T}}^L_\rho\ \interp{\ottnt{t}}^C\ x$
\item $\interp{\Gamma}^C ; \interp{\Gamma}^L \vdash \interp{\ottnt{T'}}^L\ \interp{\ottnt{t'}}^C$
\end{itemize}
\noindent We can instantiate the quantifier in the first fact, using
$\ottdrulename{Pv\_Alle}$ and Theorem~\ref{thm:sndc} (to get
$\interp{\Gamma}^C \vdash \interp{\ottnt{t'}}^C : \interp{\ottnt{T'}}^C$).  This
gives us the following from the first fact:
\[
\interp{\Gamma}^C ; \interp{\Gamma}^L \vdash \interp{\ottnt{T'}}^L_\downarrow\ x\ \Rightarrow [\interp{\ottnt{t'}}^C/x]\interp{\ottnt{T}}^L_\rho\ \interp{\ottnt{t}}^C\ \interp{\ottnt{t'}}^C
\]
\noindent The antecedent of this implication is provable from the
second fact above and $\ottkw{Terminates}\, \interp{\ottnt{t'}}^C$, giving us:
\[
\interp{\Gamma}^C ; \interp{\Gamma}^L \vdash [\interp{\ottnt{t'}}^C/x]\interp{\ottnt{T}}^L_\rho\ \interp{\ottnt{t}}^C\ \interp{\ottnt{t'}}^C
\]
\noindent Since we already have $\ottkw{Terminates}\, \interp{\ottnt{t}}^C\ \interp{\ottnt{t'}}^C$, from
this we obtain (no matter what the value of $\rho$ is) 
\[
\interp{\Gamma}^C ; \interp{\Gamma}^L \vdash [\interp{\ottnt{t'}}^C/x]\interp{\ottnt{T}}^L\ \interp{\ottnt{t}}^C\ \interp{\ottnt{t'}}^C
\]
\noindent The desired conclusion then follows from the fact that
$[\interp{\ottnt{t'}}^C/x]\interp{\ottnt{T}}^L_\rho = \interp{[ \, \ottnt{t'} \, / \, \mathit{x} \, ] \, \ottnt{T}}^L_\rho$.  We omit the straightforward proof of this fact.

Now we must consider the case where $\theta=\downarrow$, and hence
$\rho = \downarrow$ (from the rule's third premise).  In this case,
the IH gives us:
\begin{itemize}
\item $\interp{\Gamma}^C ; \interp{\Gamma}^L \vdash \ottkw{Terminates}\, \interp{\ottnt{t}}^C \wedge \forall \mathit{x} : \interp{\ottnt{T'}}^C.\ \interp{\ottnt{T'}}^L_\downarrow\ x\ \Rightarrow \interp{\ottnt{T}}^L_\downarrow\ \interp{\ottnt{t}}^C\ x$
\item $\interp{\Gamma}^C ; \interp{\Gamma}^L \vdash \ottkw{Terminates}\, \interp{\ottnt{t'}}^C \wedge \interp{\ottnt{T'}}^L\ \interp{\ottnt{t'}}^C$
\end{itemize}
\noindent We must prove:
\[
\interp{\Gamma}^C ; \interp{\Gamma}^L \vdash \interp{[ \, \ottnt{t'} \, / \, \mathit{x} \, ] \, \ottnt{T}}^L_\downarrow\ \interp{\ottnt{t} \, \ottnt{t'}}^C
\]
\noindent By the same reasoning as in the previous case, we obtain:
\[
\interp{\Gamma}^C ; \interp{\Gamma}^L \vdash \interp{[ \, \ottnt{t'} \, / \, \mathit{x} \, ] \, \ottnt{T}}^L_\rho\ \interp{\ottnt{t}}^C\ \interp{\ottnt{t'}}^C
\]
\noindent But this is exactly what we must prove, since $\rho=\downarrow$.

\pfcase{T\_Zero}{\ottdruleTXXZero{}} It suffices to prove
$\interp{\Gamma}^C ; \interp{\Gamma}^L \vdash \ottkw{Terminates}\ \, 0 \wedge
       \ottkw{True}$, which follows easily using
       $\ottdrulename{Pv\_Term0}$ and $\ottdrulename{Pv\_Truei}$
       .

\pfcase{T\_Suc}{\ottdruleTXXSuc{}} We again case-split on whether
$\theta=?$ or $\theta=\downarrow$.  In the former case, the IH gives
us:
\[
\interp{\Gamma}^C ; \interp{\Gamma}^L \vdash \ottkw{Terminates}\,\interp{\ottnt{t}}^C\wedge\ottkw{True}
\]
\noindent We must prove:
\[
\interp{\Gamma}^C ; \interp{\Gamma}^L \vdash \ottkw{Terminates}\,S\ \interp{\ottnt{t}}^C\wedge\ottkw{True}
\]
\noindent This follows by $\ottdrulename{Pv\_TerminatesS}$.  If $\theta=\downarrow$, we must prove
\[
\interp{\Gamma}^C ; \interp{\Gamma}^L \vdash \ottkw{Terminates}\,S\ \interp{\ottnt{t}}^C\Rightarrow \ottkw{True}
\]
\noindent But this is holds just by $\ottdrulename{Pv\_Impi}$ and $\ottdrulename{Pv\_Truei}$.

\pfcase{T\_Rec}{\ottdruleTXXRec{}} By the IH,
we have:
\[
\interp{\Gamma}^C ,\, f : \interp{\ottnt{T'}}^C\to\interp{\ottnt{T}}^C, x:\interp{\ottnt{T'}}^C\ ;\ 
\interp{\Gamma}^L,\, \forall x : \interp{\ottnt{T'}}^C.\ \interp{\ottnt{T'}}^L_\downarrow\ x \Rightarrow \interp{\ottnt{T}}^L_?\ (f\ x),\, \interp{\ottnt{T'}}^L_\downarrow x\ \vdash\ \interp{\ottnt{T}}^L_?\ \interp{\ottnt{t}}^C
\]
\noindent  Applying $\ottdrulename{Pv\_Impi}$ and $\ottdrulename{Pv\_Alli}$, we get:
\[
\interp{\Gamma}^C ,\, f : \interp{\ottnt{T'}}^C\to\interp{\ottnt{T}}^C\ ;\ 
\interp{\Gamma}^L,\, \forall x : \interp{\ottnt{T'}}^C.\ \interp{\ottnt{T'}}^L_\downarrow\ x \Rightarrow \interp{\ottnt{T}}^L_?\ (f\ x)\ \vdash\ \forall x:\interp{\ottnt{T'}}^C.\ \interp{\ottnt{T'}}^L_\downarrow x \Rightarrow \interp{\ottnt{T}}^L_?\ \interp{\ottnt{t}}^C
\]
\noindent This exactly matches the logical premise of the
$\ottdrulename{Pv\_CompInd}$ rule, with $F$ taken to be $\forall
x:\interp{\ottnt{T'}}^C.\ \interp{\ottnt{T'}}^L_\downarrow x \Rightarrow
\interp{\ottnt{T}}^L_?\ z$:
\[
\ottdrulePvXXCompInd{}
\]
\noindent So applying $\ottdrulename{Pv\_CompInd}$, we get the following fact (call it (J)):
\[
\interp{\Gamma}^C\ ;\ 
\interp{\Gamma}^L\ \vdash\ \forall x:\interp{\ottnt{T'}}^C.\ \ottkw{Terminates}\, (\ottkw{rec}\ f(x)\,=\, \interp{\ottnt{t}}^C)\ x \Rightarrow \interp{\ottnt{T'}}^L_\downarrow x \Rightarrow \interp{\ottnt{T}}^L_?\  (\ottkw{rec}\ f(x)\,=\, \interp{\ottnt{t}}^C)\ x
\]
\noindent Note that the consequent $\interp{\ottnt{T}}^L_?\  (\ottkw{rec}\ f(x)\,=\, \interp{\ottnt{t}}^C)\ x$ of this implication
is, by definition of $\interp{\cdot}^L_?$:
\[
\ottkw{Terminates}\ (\ottkw{rec}\ f(x)\,=\, \interp{\ottnt{t}}^C)\ x \Rightarrow \interp{\ottnt{T}}^L\  (\ottkw{rec}\ f(x)\,=\, \interp{\ottnt{t}}^C)\ x
\]
\noindent So the premise of the implication in (J) is redundant, and we can deduce the following (J'):
\[
\interp{\Gamma}^C\ ;\ 
\interp{\Gamma}^L\ \vdash\ \forall x:\interp{\ottnt{T'}}^C.\ \interp{\ottnt{T'}}^L_\downarrow x \Rightarrow \interp{\ottnt{T}}^L_?\  (\ottkw{rec}\ f(x)\,=\, \interp{\ottnt{t}}^C)\ x
\]
\noindent It suffices now to prove:
\[
\interp{\Gamma}^C ; \interp{\Gamma}^L \vdash \ottkw{Terminates}\ \ottkw{rec}\ f(x)\,=\, \interp{\ottnt{t}}^C\ \wedge
 \forall x : \interp{\ottnt{T'}}^C.\ \interp{\ottnt{T'}}^L_\downarrow\ x \Rightarrow \interp{\ottnt{T}}^L_?\ (\ottkw{rec}\ f(x)\,=\, \interp{\ottnt{t}}^C)\ x
\]
\noindent The first conjunct is provable by
$\ottdrulename{Pv\_TermRec}$.  The second now follows directly
by what we have just deduced above.

\pfcase{T\_RecNat}{\ottdruleTXXRecNat{}} By the IH,
we have
\[
\begin{array}{l}
\interp{\Gamma}^C, f : \ottkw{nat}\to\interp{\ottnt{T}}^C, x : \ottkw{nat}, p :  \ottkw{nat}  \to  ( \,  \ottkw{nat}  \to  \ottkw{nat}  \, ) \ ;\\
 \interp{\Gamma}^L, \ottkw{Terminates}\ \, \mathit{f}\ \wedge\ \ottnt{F_{{\mathrm{1}}}}, \ottkw{Terminates}\ \, \mathit{x}\ \wedge\ \ottkw{True},  \ottkw{Terminates}\ \, \mathit{p}\ \wedge\ \ottnt{F_{{\mathrm{2}}}} \ 
\vdash\ \interp{\ottnt{T}}^L_\downarrow\ \interp{\ottnt{t}}^C
\end{array}
\]
\noindent where:
\begin{eqnarray*}
\ottnt{F_{{\mathrm{1}}}} & = & \forall \mathit{x_{{\mathrm{1}}}} : \ottkw{nat}.\, \ottkw{Terminates}\ \, \mathit{x_{{\mathrm{1}}}}\ \wedge\ \ottkw{True}\ \Rightarrow\ \interp{\ottnt{T}}^L_?\ (f \mathit{x_{{\mathrm{1}}}})\\
\ottnt{F_{{\mathrm{2}}}} & = & \forall \mathit{x_{{\mathrm{1}}}} : \ottkw{nat}.\, \ottkw{Terminates}\ \, \mathit{x_{{\mathrm{1}}}}\ \wedge\ \ottkw{True}\ \Rightarrow\ \ottkw{Terminates}\ \, ( \, \mathit{p} \, \mathit{x_{{\mathrm{1}}}} \, )\ \wedge \\
\ &\ & \forall \mathit{x_{{\mathrm{2}}}} : \ottkw{nat}.\, \ottkw{Terminates}\ \, \mathit{x_{{\mathrm{2}}}}\ \wedge\ \mathit{x} \, = \, \ottkw{Suc} \, \mathit{x_{{\mathrm{1}}}}\ \Rightarrow  \ottkw{Terminates}\ \, ( \, ( \, \mathit{p} \, \mathit{x_{{\mathrm{1}}}} \, ) \, \mathit{x_{{\mathrm{2}}}} \, )\ \wedge   \\
\ &\ & \ottkw{Terminates}\ \, ( \, \mathit{f} \, \mathit{x_{{\mathrm{1}}}} \, )
\end{eqnarray*}
\noindent We may easily show that we can replace $\ottnt{F_{{\mathrm{1}}}}$ and $\ottnt{F_{{\mathrm{2}}}}$ by the following simplified versions:
\begin{eqnarray*}
\ottnt{F'_{{\mathrm{1}}}} & = & \forall \mathit{x_{{\mathrm{1}}}} : \ottkw{nat}.\, \ottkw{Terminates}\ \, \mathit{x_{{\mathrm{1}}}}\ \Rightarrow\ \interp{\ottnt{T}}^L_?\ (f \mathit{x_{{\mathrm{1}}}})\\
\ottnt{F'_{{\mathrm{2}}}} & = & \forall \mathit{x_{{\mathrm{1}}}} : \ottkw{nat}.\, \ottkw{Terminates}\ \, \mathit{x_{{\mathrm{1}}}}\ \Rightarrow\ \mathit{x} \, = \, \ottkw{Suc} \, \mathit{x_{{\mathrm{1}}}}\ \Rightarrow\ \ottkw{Terminates}\ \, ( \, \mathit{f} \, \mathit{x_{{\mathrm{1}}}} \, )
\end{eqnarray*}
\noindent This (and similar simplifications), followed by some uses of
$\ottdrulename{Pv\_Alli}$ and $\ottdrulename{Pv\_Impi}$, and also
supplying an arbitrary lambda-abstraction of type $ \ottkw{nat}  \to  ( \,  \ottkw{nat}  \to  \ottkw{nat}  \, ) $ for $p$ gives us the following central assumption (call it
(J)) from the judgment above:
\[
\interp{\Gamma}^C ; \interp{\Gamma}^L\ \vdash\ \forall x : \ottkw{nat} . \ottkw{Terminates}\ \, \mathit{x} \ \Rightarrow\ \forall f : \ottkw{nat}\to\interp{\ottnt{T}}^C.\ (\ottkw{Terminates}\ \, \mathit{f}\ \wedge\ \ottnt{F'_{{\mathrm{1}}}}\ \wedge\ \ottnt{F'_{{\mathrm{2}}}})\ \Rightarrow\ \interp{\ottnt{T}}^L_\downarrow\ \interp{\ottnt{t}}^C
\]
\noindent It suffices to show:
\[
\interp{\Gamma}^C ; \interp{\Gamma}^L \vdash \ottkw{Terminates}\ \ottkw{rec}\,f(x)\,=\interp{\ottnt{t}}^C\ \wedge\ \forall x : \ottkw{nat}.\, \ottkw{Terminates}\ \, \mathit{x}\ \wedge\ \ottkw{True}\ \Rightarrow\ \interp{\ottnt{T}}^L_\downarrow\ (\ottkw{rec}\,f(x)\,=\interp{\ottnt{t}}^C)\ x
\]
\noindent We have the first conjunct by $\ottdrulename{Pv\_TermRec}$.  For the second,
it suffices to show:
\[
\interp{\Gamma}^C ; \interp{\Gamma}^L \vdash \forall x : \ottkw{nat}.\, \ottkw{Terminates}\ \, \mathit{x}\ \Rightarrow\ \interp{\ottnt{T}}^L_\downarrow\ (\ottkw{rec}\,f(x)\,=\interp{\ottnt{t}}^C)\ x
\]
\noindent We do this by induction (using $\ottdrulename{Pv\_Ind}$).
First, though, we observe that the reasoning we used in the previous
case (for $\ottdrulename{T\_Rec}$) to prove what we called (J')
applies here (except that here we have some additional assumptions in
the context).  This lets us deduce the following, which we will call
(J') (essentially the (J') from the previous case, with $\ottnt{T'}$
replaced by $\ottkw{nat}$):
\[
\forall \mathit{x} : \ottkw{nat}.\, \ottkw{Terminates}\ \, \mathit{x}\ \Rightarrow\ \interp{\ottnt{T}}^L_?\ (\ottkw{rec}\ f(x)\,=\, \interp{\ottnt{t}}^C)\ x
\] 
\noindent So now for the base case, we must prove:
\[
\interp{\Gamma}^C ; \interp{\Gamma}^L \vdash \interp{[ \, 0 \, / \, \mathit{x} \, ] \, \ottnt{T}}^L_\downarrow\ (\ottkw{rec}\,f(x)\,=\interp{\ottnt{t}}^C)\ 0
\]
\noindent This follows easily using $\ottdrulename{Pv\_Subst}$ and $\ottdrulename{Pv\_OpSem}$ from:
\[
\interp{\Gamma}^C ; \interp{\Gamma}^L \vdash \interp{[ \, 0 \, / \, \mathit{x} \, ] \, \ottnt{T}}^L_\downarrow\ ([ 0 / x ] [ \ottkw{rec}\,f(x)\,=\interp{\ottnt{t}}^C / f] \interp{\ottnt{t}}^C
\]
\noindent We obtain this by instantiating (J) above with $0$ for
$\mathit{x}$, and $\ottkw{rec}\,f(x)\,=\interp{\ottnt{t}}^C$ for $f$.  We have
the required proofs of termination by $\ottdrulename{Pv\_Term0}$
and $\ottdrulename{Pv\_TermRec}$.  We have a proof of the
appropriately instantiated premise $\ottnt{F'_{{\mathrm{1}}}}$ of (J), since this is
exactly (J').  We easily prove the instantiation of premise $\ottnt{F'_{{\mathrm{2}}}}$
of (J), since this is:
\[
\forall \mathit{x_{{\mathrm{1}}}} : \ottkw{nat}.\, \ottkw{Terminates}\ \, \mathit{x_{{\mathrm{1}}}}\ \Rightarrow\ 0 \, = \, \ottkw{Suc} \, \mathit{x_{{\mathrm{1}}}}\ \Rightarrow\ \ottkw{Terminates}\ ((\ottkw{rec}\,f(x)\,=\interp{\ottnt{t}}^C) \mathit{x_{{\mathrm{1}}}})
\]
\noindent This formula is easily proved using
$\ottdrulename{Pv\_Contra}$ with premise $0 \, = \, \ottkw{Suc} \, \mathit{x_{{\mathrm{1}}}}$.  So from (J),
with these instantiations and proven premises, we obtain the following,
which is exactly what we had to prove:
\[
\interp{\Gamma}^C ; \interp{\Gamma}^L \vdash \interp{[ \, 0 \, / \, \mathit{x} \, ] \, \ottnt{T}}^L_\downarrow\ (\ottkw{rec}\,f(x)\,=\interp{\ottnt{t}}^C)\ 0
\]
\noindent For the step case, we must now prove:
\[
\interp{\Gamma}^C,\, x':\ottkw{nat} ; \interp{\Gamma}^L,\, \ottkw{Terminates}\ \, \mathit{x'},\, \interp{[ \, \mathit{x'} \, / \, \mathit{x} \, ] \, \ottnt{T}}^L_\downarrow\ (\ottkw{rec}\,f(x)\,=\interp{\ottnt{t}}^C)\ x'\ \vdash\ \interp{[ \, \ottkw{Suc} \, \mathit{x'} \, / \, \mathit{x} \, ] \, \ottnt{T}}^L_\downarrow\ (\ottkw{rec}\,f(x)\,=\interp{\ottnt{t}}^C)\ (\ottkw{S}\ x')
\]
\noindent Now we instantiate (J) above with $\ottkw{Suc} \, \mathit{x'}$ for $x$, and
again $\ottkw{rec}\,f(x)\,=\interp{\ottnt{t}}^C$ for $f$.  We easily obtain
the required proofs of termination.  The instantiated $\ottnt{F'_{{\mathrm{1}}}}$ we again
have by (J').  The instantiated premise $\ottnt{F'_{{\mathrm{2}}}}$ is:
\[
\forall \mathit{x_{{\mathrm{1}}}} : \ottkw{nat}.\, \ottkw{Terminates}\ \, \mathit{x_{{\mathrm{1}}}}\ \Rightarrow\ \ottkw{Suc} \, \mathit{x'} \, = \, \ottkw{Suc} \, \mathit{x_{{\mathrm{1}}}}\ \Rightarrow\ \ottkw{Terminates}\ ((\ottkw{rec}\,f(x)\,=\interp{\ottnt{t}}^C) \mathit{x_{{\mathrm{1}}}})
\]
\noindent We can prove this premise as follows.  Assume arbitrary
terminating $\mathit{x_{{\mathrm{1}}}}$ of sort $\ottkw{nat}$, and assume $\ottkw{Suc} \, \mathit{x'} \, = \, \ottkw{Suc} \, \mathit{x_{{\mathrm{1}}}}$.
Using $\ottdrulename{Pv\_Subst}$ and $\ottdrulename{Pv\_OpSem}$, we can
derive $\mathit{x'} \, = \, \mathit{x_{{\mathrm{1}}}}$ from this:
\[
\infer[\ottdrulename{Pv\_Subst}]
  {\mathit{x'} \, = \, \mathit{x_{{\mathrm{1}}}}}
  {\infer[\ottdrulename{Pv\_Subst}]
         { \ottkw{case} \  ( \, \ottkw{Suc} \, \mathit{x_{{\mathrm{1}}}} \, ) \  0 \  \lambda \, \mathit{z} \, . \, \mathit{z}  \, = \, \mathit{x'}}{\ottkw{Suc} \, \mathit{x'} \, = \, \ottkw{Suc} \, \mathit{x_{{\mathrm{1}}}} & \infer[\ottdrulename{Pv\_OpSem}]{ \ottkw{case} \  ( \, \ottkw{Suc} \, \mathit{x'} \, ) \  0 \  \lambda \, \mathit{z} \, . \, \mathit{z}  = \mathit{x'}}{\ }}
  & \infer[\ottdrulename{Pv\_OpSem}]{ \ottkw{case} \  ( \, \ottkw{Suc} \, \mathit{x_{{\mathrm{1}}}} \, ) \  0 \  \lambda \, \mathit{z} \, . \, \mathit{z}  \, = \, \mathit{x_{{\mathrm{1}}}}}{\ }}
\]
\noindent So now to complete the proof of the instantiated premise $\ottnt{F'_{{\mathrm{2}}}}$, we need only prove 
\[
\ottkw{Terminates}\ (\ottkw{rec}\,f(x)\,=\interp{\ottnt{t}}^C)\ x'
\]
\noindent But this follows directly from the assumption we have in this step case of $\ottdrulename{Pv\_Ind}$:
\[
\interp{[ \, \mathit{x'} \, / \, \mathit{x} \, ] \, \ottnt{T}}^L_\downarrow\ (\ottkw{rec}\,f(x)\,=\interp{\ottnt{t}}^C)\ x'
\]
\noindent So we have all the premises required by (J), and we can
prove the following (applying a derived weakening rule, whose easy
proof is omitted, to (J) to add our other assumptions to its
contexts):
\[
\interp{[ \, \ottkw{Suc} \, \mathit{x'} \, / \, \mathit{x} \, ] \, \ottnt{T}}^L_\downarrow\ [ \ottkw{Suc} \, \mathit{x'} / x ] [ (\ottkw{rec}\,f(x)\,=\interp{\ottnt{t}}^C) / f ]\interp{\ottnt{t}}^C
\]
\noindent This now easily implies the required conclusion, using
$\ottdrulename{Pv\_Subst}$ with the following equation, which holds
by $\ottdrulename{Pv\_OpSem}$:
\[
[ \ottkw{Suc} \, \mathit{x'} / x ] [ (\ottkw{rec}\,f(x)\,=\interp{\ottnt{t}}^C) / f ]\interp{\ottnt{t}}^C = (\ottkw{rec}\,f(x)\,=\interp{\ottnt{t}}^C) (\ottkw{Suc} \, \mathit{x'})
\]

\pfcase{T\_Case}{\ottdruleTXXCase{}} As for some cases above, we begin by
case-splitting on whether $\theta=?$ or $\theta=\downarrow$.  Suppose
$\theta=?$.  Then applying the IH to the second and third premises,
and then a few basic logical simplifications, gives us:
\begin{enumerate}
\item $\interp{\Gamma}^C ; \interp{\Gamma}^L \vdash \ottkw{Terminates}\,\interp{\ottnt{t'}}^C \Rightarrow \interp{[ \, 0 \, / \, \mathit{x} \, ] \, \ottnt{T}}^L \interp{\ottnt{t'}}^C$
\item $\interp{\Gamma}^C ; \interp{\Gamma}^L \vdash \ottkw{Terminates}\,\interp{\ottnt{t''}}^C \Rightarrow \forall x' : \ottkw{nat}. \ottkw{Terminates}\, x' \Rightarrow \interp{[ \, \ottkw{Suc} \, \mathit{x'} \, / \, \mathit{x} \, ] \, \ottnt{T}}^L_\rho\ (\interp{\ottnt{t''}}^C\ x')$
\end{enumerate}
\noindent We must show:
\[
\interp{\Gamma}^C ; \interp{\Gamma}^L \vdash \ottkw{Terminates}\,\interp{ \ottkw{case} \  \ottnt{t} \  \ottnt{t'} \  \ottnt{t''} }^C \Rightarrow \interp{[ \, \ottnt{t} \, / \, \mathit{x} \, ] \, \ottnt{T}}^L \interp{ \ottkw{case} \  \ottnt{t} \  \ottnt{t'} \  \ottnt{t''} }^C
\]
\noindent So assume $\ottkw{Terminates}\,\interp{ \ottkw{case} \  \ottnt{t} \  \ottnt{t'} \  \ottnt{t''} }^C$, and show $\interp{[ \, \ottnt{t} \, / \, \mathit{x} \, ] \, \ottnt{T}}^L \interp{ \ottkw{case} \  \ottnt{t} \  \ottnt{t'} \  \ottnt{t''} }^C$.  By $\ottdrulename{Pv\_TermInv}$, this
assumption implies $\ottkw{Terminates}\,\interp{\ottnt{t}}^C$.  Since
$\interp{\Gamma}^C \vdash \interp{\ottnt{t}}^C:\ottkw{nat}$ by Theorem~\ref{thm:sndc},
we will now seek to prove the following by $\ottdrulename{Pv\_Ind}$:
\[
\interp{\Gamma}^C ; \interp{\Gamma}^L \vdash \forall x : \ottkw{nat} . \ottkw{Terminates}\ \, \mathit{x} \Rightarrow \ottkw{Terminates}\,\interp{ \ottkw{case} \  \mathit{x} \  \ottnt{t'} \  \ottnt{t''} }^C \Rightarrow \interp{\ottnt{T}}^L \interp{ \ottkw{case} \  \mathit{x} \  \ottnt{t'} \  \ottnt{t''} }^C
\]
\noindent If we can derive this judgment, then we can instantiate
$\mathit{x}$ with $\interp{\ottnt{t}}^C$ (for which we have
$\ottkw{Terminates}\,\interp{\ottnt{t}}^C$) to conclude the desired
\[
\interp{\Gamma}^C ; \interp{\Gamma}^L \vdash \ottkw{Terminates}\,\interp{ \ottkw{case} \  \ottnt{t} \  \ottnt{t'} \  \ottnt{t''} }^C \Rightarrow \interp{[ \, \ottnt{t} \, / \, \mathit{x} \, ] \, \ottnt{T}}^L \interp{ \ottkw{case} \  \ottnt{t} \  \ottnt{t'} \  \ottnt{t''} }^C
\]
\noindent To apply $\ottdrulename{Pv\_Ind}$ as desired, we must prove the base case and step case of the induction:
\begin{itemize}
\item $\interp{\Gamma}^C ; \interp{\Gamma}^L \vdash \ottkw{Terminates}\,\interp{ \ottkw{case} \  0 \  \ottnt{t'} \  \ottnt{t''} }^C \Rightarrow \interp{[ \, 0 \, / \, \mathit{x} \, ] \, \ottnt{T}}^L \interp{ \ottkw{case} \  0 \  \ottnt{t'} \  \ottnt{t''} }^C$
\item $\interp{\Gamma}^C, x' : \ottkw{nat} ; \interp{\Gamma}^L , \ottkw{Terminates}\ \, \mathit{x'} \vdash \ottkw{Terminates}\,\interp{ \ottkw{case} \  ( \, \ottkw{Suc} \, \mathit{x'} \, ) \  \ottnt{t'} \  \ottnt{t''} }^C \Rightarrow \interp{[ \, \ottkw{Suc} \, \mathit{x'} \, / \, \mathit{x} \, ] \, \ottnt{T}}^L \interp{ \ottkw{case} \  ( \, \ottkw{Suc} \, \mathit{x'} \, ) \  \ottnt{t'} \  \ottnt{t''} }^C$
\end{itemize}
\noindent This base case follows from fact (1) above, using the
equation $\interp{ \ottkw{case} \  0 \  \ottnt{t'} \  \ottnt{t''} }^C = \interp{\ottnt{t'}}^C$.  This
equation is easily shown by the definition of $\interp{\cdot}^C$ and
$\ottdrulename{Pv\_OpSem}$.  So we now prove the step case.  First, we
simplify the desired formula using the easily proved equation
$\interp{ \ottkw{case} \  ( \, \ottkw{Suc} \, \mathit{x'} \, ) \  \ottnt{t'} \  \ottnt{t''} }^C = \interp{\ottnt{t''} \, \mathit{x'}}^C$.  So our
new goal formula is
\[
\interp{\Gamma}^C, x' : \ottkw{nat} ; \interp{\Gamma}^L , \ottkw{Terminates}\ \, \mathit{x'} \vdash \ottkw{Terminates}\,\interp{\ottnt{t''} \, \mathit{x'}}^C \Rightarrow \interp{[ \, \ottkw{Suc} \, \mathit{x'} \, / \, \mathit{x} \, ] \, \ottnt{T}}^L \interp{\ottnt{t''} \, \mathit{x'}}^C
\]
\noindent So assume $\ottkw{Terminates}\ \, \mathit{x'}$ and
$\ottkw{Terminates}\,\interp{\ottnt{t''} \, \mathit{x'}}^C$, and show $\interp{[ \, \ottkw{Suc} \, \mathit{x'} \, / \, \mathit{x} \, ] \, \ottnt{T}}^L \interp{\ottnt{t''} \, \mathit{x'}}^C$.  Instantiating fact
(2) above with $\mathit{x'}$ and these assumptions, we get:
\[
\interp{[ \, \ottkw{Suc} \, \mathit{x'} \, / \, \mathit{x} \, ] \, \ottnt{T}}^L_\rho\ (\interp{\ottnt{t''}}^C\ x')
\]
\noindent This implies the desired formula in either possible case for
$\rho$.

Now let us assume $\theta=\downarrow$.  This case is similar to the
above, so we give fewer details.  The IH for the three premises gives
us:
\begin{enumerate}
\item $\interp{\Gamma}^C ; \interp{\Gamma}^L \vdash \ottkw{Terminates}\,\interp{\ottnt{t}}^C \wedge \ottkw{True}$
\item $\interp{\Gamma}^C ; \interp{\Gamma}^L \vdash \ottkw{Terminates}\,\interp{\ottnt{t'}}^C \wedge \interp{[ \, 0 \, / \, \mathit{x} \, ] \, \ottnt{T}}^L \interp{\ottnt{t'}}^C$
\item $\interp{\Gamma}^C ; \interp{\Gamma}^L \vdash \ottkw{Terminates}\,\interp{\ottnt{t''}}^C \wedge \forall x' : \ottkw{nat}. \ottkw{Terminates}\, x' \Rightarrow \interp{[ \, \ottkw{Suc} \, \mathit{x'} \, / \, \mathit{x} \, ] \, \ottnt{T}}^L_\rho\ (\interp{\ottnt{t''}}^C\ x')$
\end{enumerate}
\noindent We must show
\[
\interp{\Gamma}^C ; \interp{\Gamma}^L \vdash \ottkw{Terminates}\,\interp{ \ottkw{case} \  \ottnt{t} \  \ottnt{t'} \  \ottnt{t''} }^C \wedge \interp{[ \, \ottnt{t} \, / \, \mathit{x} \, ] \, \ottnt{T}}^L \interp{ \ottkw{case} \  \ottnt{t} \  \ottnt{t'} \  \ottnt{t''} }^C
\]
\noindent Since we have $\ottkw{Terminates}\ \interp{\ottnt{t}}^C$ and
$\interp{\Gamma}^C\vdash \interp{\ottnt{t}}^C:\ottkw{nat}$, it suffices to
prove the following more general statement:
\[
\interp{\Gamma}^C ; \interp{\Gamma}^L \vdash \forall x : \ottkw{nat} . \ottkw{Terminates}\ \, \mathit{x} \Rightarrow \ottkw{Terminates}\,\interp{ \ottkw{case} \  \mathit{x} \  \ottnt{t'} \  \ottnt{t''} }^C \wedge \interp{\ottnt{T}}^L \interp{ \ottkw{case} \  \mathit{x} \  \ottnt{t'} \  \ottnt{t''} }^C
\]
\noindent We again apply $\ottdrulename{Pv\_Ind}$.  The base case is
again immediate using $\interp{ \ottkw{case} \  0 \  \ottnt{t'} \  \ottnt{t''} }^C =
\interp{\ottnt{t'}}^C$.  Similarly reasoning as for the step case above
gives us:
\[
\interp{[ \, \ottkw{Suc} \, \mathit{x'} \, / \, \mathit{x} \, ] \, \ottnt{T}}^L_\rho\ (\interp{\ottnt{t''}}^C\ x')
\]
\noindent Since $\rho$ must be $\downarrow$ in this case, we obtain
from this fact the desired $\ottkw{Terminates}\ (\interp{\ottnt{t''}}^C\ x')$,
as well as
\[
\interp{[ \, \ottkw{Suc} \, \mathit{x'} \, / \, \mathit{x} \, ] \, \ottnt{T}}^L\ (\interp{\ottnt{t''}}^C\ x')
\]

\pfcase{T\_Contra}{\ottdruleTXXContra{}} By the IH we have $\interp{\Gamma}^C ;
\interp{\Gamma}^L \vdash \ottkw{Terminates}\ \interp{\ottnt{t}}^C\ \wedge\ 0 =
\ottkw{S}\ \interp{\ottnt{t'}}^C$.  We must show $\interp{\Gamma}^C ;
\interp{\Gamma}^L \vdash \interp{\ottnt{T}}^L_\theta\ 0$.  But this fact
follows directly from the second conjunct of the fact we have, using
$\ottdrulename{Pv\_Contra}$.

\pfcase{T\_Abort}{\ottdruleTXXAbort{}} We must prove $\interp{\Gamma}^C ;
\interp{\Gamma}^L \vdash \ottkw{Terminates}\ \, \ottkw{abort} \Rightarrow
\interp{\ottnt{T}}^L\ \ottkw{abort}$.  But this follows directly by
$\ottdrulename{Pv\_Impi}$ from
$\ottdrulename{Pv\_NotTerminatesAbort}$.


\section{Typing rules for system $W'$}
\label{sec:wptp}
\ottdefnstassn{}











\fi 

\end{document}